\documentclass[a4paper,fleqn,usenatbib]{mnras}

\usepackage{newtxtext,newtxmath}

\usepackage[T1]{fontenc}
\usepackage{ae,aecompl}

\usepackage{graphicx}	\usepackage{amsmath}	\usepackage{amssymb}	\usepackage{caption}
\usepackage{subcaption}
\usepackage{hyperref}

\usepackage[english]{babel}
\usepackage[autostyle]{csquotes}
\MakeOuterQuote{"}

\usepackage{tabularx}

\newcommand{\kms}{\,kms$^{-1}$} \newcommand{\NaI}{NaI$_{\textrm{SDSS}}$}
\newcommand{\ppxf}{\texttt{pPXF}}

\title[Radial measurements of far-red indices]{Radial measurements of IMF-sensitive absorption features in two massive ETGs}

\author[S. P. Vaughan et al.]{Sam P. Vaughan$^{1}$\thanks{E-mail: sam.vaughan@physics.ox.ac.uk (SPV)},  Roger L. Davies$^{1}$, Simon Zieleniewski$^{1}$
\newauthor
and Ryan C. W. Houghton$^{1}$\\
$^{1}$Department of Astrophysics, University of Oxford, Denys Wilkinson Building, Keble Road, Oxford, OX1 4RH\\
}

\date{Accepted 2017 December 7. Received 2017 December 2017; in original form 2016 December 1}

\pubyear{2017}

\begin{document}
\label{firstpage}
\pagerange{\pageref{firstpage}--\pageref{lastpage}}
\maketitle

\begin{abstract}
We make radial measurements of stellar initial mass function (IMF) sensitive absorption features in the two massive early-type galaxies NGC~1277 and IC~843. Using the Oxford Short Wavelength Integral Field SpecTrogaph (SWIFT), we obtain resolved measurements of the NaI0.82 and FeH0.99 indices, among others, finding both galaxies show strong gradients in NaI absorption combined with flat FeH profiles at $\sim0.4$\AA. We find these measurements may be explained by radial gradients in the IMF, appropriate abundance gradients in [Na/Fe] and [Fe/H], or a combination of the two, and our data is unable to break this degeneracy. We also use full spectral fitting to infer global properties from an integrated spectrum of each object, deriving a unimodal IMF slope consistent with Salpeter in IC~843 ($x=2.27\pm0.17$) but steeper than Salpeter in NGC~1277 ($x=2.69\pm0.11$), despite their similar FeH equivalent widths. Independently, we fit the strength of the FeH feature and compare to the E-MILES and CvD12 stellar population libraries, finding agreement between the models. The IMF values derived in this way are in close agreement with those from spectral fitting in NGC~1277 ($x_{\mathrm{CvD}}=2.59^{+0.25}_{-0.48}$ , $x_{\mathrm{E-MILES}}=2.77\pm0.31$), but are less consistent in IC~843, with the IMF derived from FeH alone leading to steeper slopes than when fitting the full spectrum ($x_{\mathrm{CvD}}=2.57^{+0.30}_{-0.41}$, $x_{\mathrm{E-MILES}}=2.72\pm0.25$). This work highlights the importance of a large wavelength coverage for breaking the degeneracy between abundance and IMF variations, and may bring into doubt the use of the Wing-Ford band as an IMF index if used without other spectral information. 
\end{abstract}


\section{Introduction}
The stellar Initial Mass Function (IMF) is of fundamental importance for understanding the evolution and present day stellar content of galaxies. The IMF defines the number density of stars at each mass on the zero age main sequence in a population, and is thus intricately linked to the small-scale, turbulent and not-well-understood process of star formation whilst also defining global properties for the population as a whole. The low mass end of the IMF, and hence the number of low mass stars, greatly affects the mass-to-light ratio $(M/L)$ of a system, since a large proportion of the stellar mass in a galaxy comes from stars below 1 M$_{\odot}$. The fact that these low mass stars contribute so little to the integrated light of a population means that large changes in the $M/L$ will not necessarily be reflected in large changes to the properties of the light itself.  The high mass slope of the IMF makes a contribution to a galaxy's $M/L$ ratio too, via stellar remnants, and also defines the importance of stellar feedback and the amount of chemical enrichment that takes place. A form for the IMF is assumed whenever a stellar mass or star formation rate is calculated, and the implications for such observational parameters if the IMF is not universal could be very serious 
\citep[e.g][]{2016MNRAS.462.2832C}.

Early efforts to measure the IMF were pioneered by \citet{Salpeter}, who used direct star counts to parametrize the IMF as a power law of the form $\xi(m)=km^{-x}$ with an exponent of $x=2.35$. Using a single power law to describe the IMF has come to be called a "unimodal" description. The value of the Salpeter exponent at the high mass end has remained remarkably constant in the numerous studies of our own galaxy since, with modern day IMF parameterisations of the Milky Way incorporating a flattening at low masses: e.g  \citet{Kroupa}  and \citet{Chabrier}.  An IMF with a power law at masses greater than 0.6 $M_{\odot}$, a flat low-mass end and a spline interpolation linking the two regimes is described as a "bimodal" IMF \citep{1996ApJS..106..307V}.  The high end slope of a bimodal IMF is defined by a power law index $\Gamma_{b}$, which is related to $x$ via $x=\Gamma_{b}+1$.  An increase in $\Gamma_{b}$, like an increase in $x$, implies an increase in the dwarf-to-giant ratio and therefore an increase in the number of low mass stars.

Historically, little evidence was found for an IMF in our galaxy which varied depending on parameters such as metallicity or environment (see \citet{Bastian} for a review). More recently, however, evidence has emerged for a non-universal IMF in studies of the unresolved stellar populations of ETGs. Dynamical modelling of galaxy kinematics undertaken by the ATLAS3D team \citep{2011MNRAS.413..813C} and \cite{2011MNRAS.415..545T} have shown that the M/L ratios of ETGs compared to the M/L ratio for a population with a Salpeter IMF diverge systematically with velocity dispersion, implying that more massive ETGs have "heavier" IMFs \citep[e.g.][]{2013MNRAS.432.1862C}. Such a dynamical analysis cannot determine whether these IMFs are "bottom heavy" (more dwarf stars) or "top heavy" (more stellar remnants), however.  Comparisons between stellar population synthesis models and strong gravitational lensing predict a similar IMF-$\sigma$ relation \citep[e.g.][]{2010ApJ...709.1195T}, although massive ETGs with Milky Way-like normalization have also been found \citep[e.g.][]{2015MNRAS.449.3441S}. 

This work concerns a third method of studying the IMF in extragalactic objects. Certain absorption features in the spectra of integrated stellar populations vary in strength between (otherwise identical) low mass dwarf stars and low mass giants. A measurement of the strength of these "gravity sensitive" indices gives a direct handle on the dwarf-to-giant ratio in a population, and hence the low-mass IMF slope. Important far red gravity sensitive absorption features include the sodium NaI doublet at 8190 \AA~\citep{1971ApJS...22..445S, 1980LicOB.823....1F, 1997ApJ...479..902S}, the calcium triplet \citep[CaT:][]{Cenarro2001} and Iron Hydride or the "Wing-Ford band" at 9916 \AA~\citep[FeH:][]{1969PASP...81..527W, 1997ApJ...484..499S}. Studying gravity sensitive absorption features in the spectra of ETGs in this way has a long history \citep[e.g.][]{1971ApJS...22..445S, 1978ApJ...221..788C, 1980LicOB.823....1F, Couture_Hardy, 2003MNRAS.339L..12C}, before more recent work by \citet{2010Natur.468..940V, vDC12} reignited interest in the topic. 

Studies of optical and far red spectral lines have suggested correlations between the IMF and [Mg/Fe] \citep{CvD12}, metallicity \citep{MartinNavarro2015c}, total dynamical density \citep{Spiniello2015a} and central velocity dispersion \citep{LaBarbera2013}, but importantly the agreement between spectral and dynamical IMF determination is unclear. \citet{Smith2014} compared the IMF slopes derived using spectroscopic methods in \citet{CvD12} and dynamical methods in \citet{2013MNRAS.432.1862C} for galaxies in common between the two studies. He found overall agreement between the two methods regarding the overarching trends presented in each study, but no correlation at all between the IMF slopes determined by each group on a galaxy by galaxy basis.   On the other hand, assuming a bimodal IMF parameterisation, \cite{2016MNRAS.463.3220L} do find agreement between spectroscopic and dynamical techniques in the central regions of 27 galaxies in the CALIFA survey. Additional investigation of individual galaxies using independent IMF measurements, rather than comparison of global trends between populations, is required to understand and explain this disagreement. 

A more technically challenging goal in spectral IMF measurements is determining whether IMF gradients exist within a single object. Formation pathways of ETGs predict "inside-out growth", where the centre of a massive galaxy forms in a single starburst event before minor mergers with satellites accrete matter at larger radii \citep[e.g.][and references therein]{2009ApJ...699L.178N, 2009MNRAS.398..898H}. IMF gradients can naturally arise from such a formation history if the global IMF differs between merger pairs, but few studies have presented evidence for such gradients to date. \cite{LaBarbera2016} measure an IMF gradient in a massive ETG with central $\sigma\sim300$\kms, whilst \cite{2015MNRAS.447.1033M} report IMF gradients in two nearby ETGs. \citet[hereafter MN15]{MartinNavarro}  also find a mild gradient in a bimodal IMF in NGC~1277, one of the objects studied in this work. Other studies make radial measurements of gravity sensitive indices but conclude in favour of individual elemental abundance gradients rather than a change in the IMF: see \citet{Z15}, \cite{2017MNRAS.465..192Z} and \citet{McConnell2016}. 

In this work, we present radial observations of gravity sensitive absorption features in two galaxies. The first, NGC~1277, is a massive, compact ETG located in the Perseus cluster ($z=0.01704$). NGC~1277 is a well studied object. It has been named as a candidate "relic galaxy" due to its similarity with ETGs at much higher redshifts \citep{2014ApJ...780L..20T}, seen controversy over the mass of its central black hole (e.g. see \citealp{2012Natur.491..729V} compared to \citealp{2013MNRAS.433.1862E}) and had radial measurements of its IMF gradient taken, found using optical and far red absorption indices (MN15). Their study didn't extend to measurements of the FeH index, however. MN15 found a bottom heavy bimodal IMF at all radii, measuring the slope of the IMF to be  $\Gamma_{b} \sim 3$ (the same high-mass slope as a unimodal power law with $x=4$) in the central regions and dropping to $\Gamma_{b} \sim 2.5$ ($x=3.5$) at radii greater than 0.6 $R_{e}$. 

The second galaxy, IC~843, is an edge on ETG located on the edge of the Coma cluster ($z=0.02457$). \cite{Thomas} conducted a study of the dark matter content of 17 ETGs in Coma, finding that IC~843 had an unusually high mass-to-light ratio in the $R_{c}$ band with the best fitting model implying that mass follows light in this system. This result could be explained by a bottom heavy IMF, but also by a dark matter distribution where the dark matter closely follows the visible matter. Both galaxies were chosen because the evidence for their heavy IMFs implies that the Wing-Ford band could be particularly strong in these objects.

This paper is organised as follows. Section \ref{sec:Obs} summarises our observations describes the data reduction process, including details of sky subtraction and telluric correction. We summarise our radial index measurements in section \ref{sec:results}, present our interpretations in section \ref{sec:Discussion} and draw our conclusions in section \ref{sec:Conclusion}. Appendices contain further discussion of our telluric correction and sky subtraction techniques.  We adopt a $\Lambda$CDM cosmology, with H$_{0}$=68\kms, $\Omega_{m}$=0.3 and $\Omega_{\Lambda}$=0.7.

\begin{table*}
	\centering
	\caption{Targets and Observations}
	\label{tab:observations}
	\begin{tabular}{lccccccc} 		\hline
		\hline

		Galaxy & D & Ra & Dec & z & R$_{e}$ & Obs. Date &  Integration Time \\
		& (Mpc) & & &  &(kpc) & & (s)  \\

		\hline
		\\

		NGC~1277 & 74.4 & 03:19:51.5 & +41:34:24.3 & 0.01704 & 1.2 & 27th Jan 2016  &  7 $\times$~900 \\
		IC~843 & 107.9 & 13:01:33.6 & +29:07:49.7 & 0.02457 & 4.7 & 17th Mar 2016  &  9 $\times$~900 \\
		 \\

		\hline
	\end{tabular}
\end{table*}

\section{Observations and Data Reduction}
\label{sec:Obs}

We used the Short Wavelength Integral Field specTrograph \citep[SWIFT]{2006SPIE.6269E..3LT} on January 27th 2016 and March 17th 2016 to obtain deep integral field observations of NGC~1277 and IC~843. Observations were taken in the 235 mas spaxel$^{-1}$ settings, giving a field-of-view of 10.3\arcsec~by 20.9\arcsec. The wavelength coverages extends from 6300 \AA~to 10412 \AA, with an average spectral resolution of R$\sim$4000 and a sampling of 1 \AA~pix$^{-1}$.  Dedicated sky frames, offset by $\sim$100\arcsec~in declination, were observed in an OSO pattern to be used as first order sky subtraction. The seeing ranged between $\sim$1\arcsec~and 1.5\arcsec~throughout the observations. Table \ref{tab:observations} lists details of the targets and observations. 

The wavelength range of SWIFT allows for measurements of the NaI 0.82, CaII triplet, MgI 0.88, TiO 0.89a and FeH (Wing Ford band) absorption features. Definitions of pseudo-continuum and absorption bands for each index, taken from \citet{Cenarro2001} and \citet[]{CvD12a}, are given in Table \ref{tab:Index_Defs}.  We use the  \NaI~definition of the NaI 0.82 index from \citet{LaBarbera2013}.

\begin{table}
	\centering
	\caption{Definitions of the feature bandpass and blue and red pseudo-continuum bandpasses for each index studied in this work, from \protect\citet{Cenarro2001} and \protect\citet{CvD12a}. The  \NaI~definition is from \protect \citet{LaBarbera2013}.  Since it is a ratio between the blue and red pseudo-continuua, the TiO index has no feature bandpass definition. All wavelengths are measured in air.}
	\label{tab:Index_Defs}
	\begin{tabular}{lccc}
    		\hline
		Index&Blue Continuum&Feature&Red Continuum \\
		& (\AA) & (\AA) &(\AA)\\
		\hline
		\NaI &  8143.0-8153.0 & 8180.0-8200.0 & 8233.0-8244.0\\
		CaT & 8474.0-8484.0 & 8484.0-8513.0 & 8563.0-8577.0\\
		& 8474.0-8484.0 & 8522.0-8562.0 & 8563.0-8577.0\\
		& 8619.0-8642.0 & 8642.0-8682.0 & 8700.0-8725.0\\
		MgI & 8777.4-8789.4 & 8801.9-8816.9 & 8847.4-8857.4\\
		TiO & 8835.0-8855.0 & --- & 8870.0-8890.0\\
		FeH & 9855.0-9880.0 & 9905.0-9935.0 & 9940.0-9970.0\\
	\end{tabular}
\end{table}

\label{sec:data_red}

The data were reduced using the SWIFT data reduction pipeline to perform standard bias subtraction, flat-field and illumination correction, wavelength calibration and error propagation. Cosmic ray hits were detected and removed using the LaCosmic routine \citep{LaCosmic}.

Differential atmospheric refraction causes the centre of the galaxy to change position within a datacube as a function of wavelength. Although the magnitude of this effect is small (leading to a $\sim$1\arcsec~shift at red wavelengths for the observations which are lowest in the sky), individual cubes were corrected by interpolating each wavelength slice to a common position. The individual observation cubes were combined using a dedicated python script, which linearly interpolates sub-pixel offsets between the frames. 

\begin{figure}
\centering
\includegraphics[width=\linewidth]{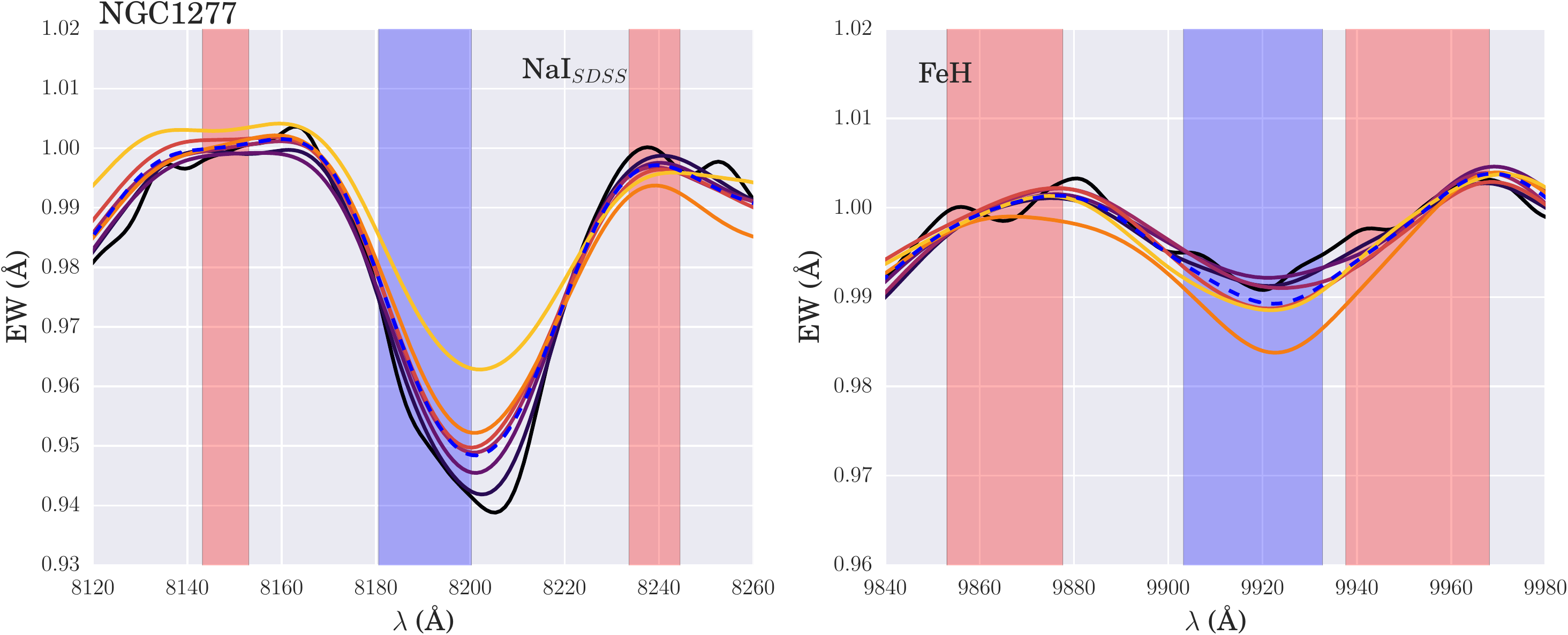}
\includegraphics[width=\linewidth]{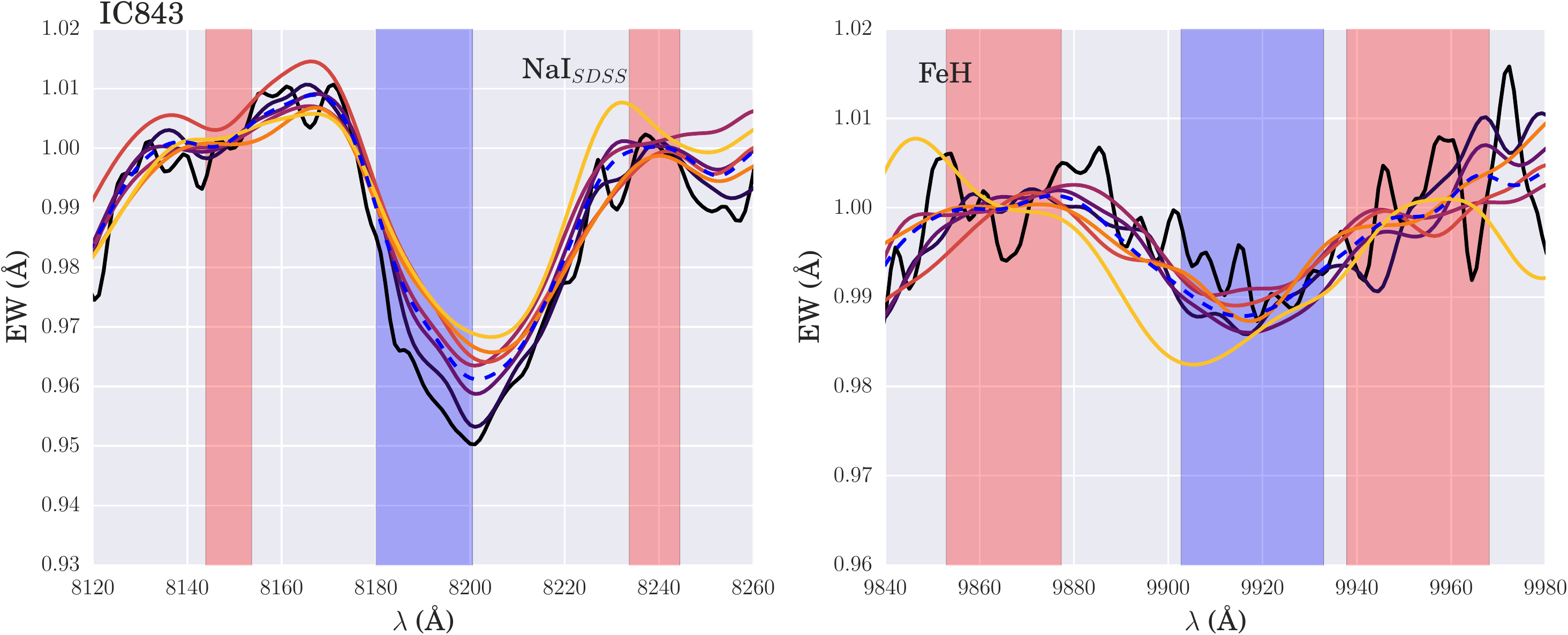}
\caption{Plots of the spectra around the IMF sensitive indices NaI0.82 and Wing-Ford band (FeH) for NGC~1277 (top) and IC~843 (bottom). Spectra are coloured from dark (central bin) to light (outskirts) and are convolved up to a common velocity dispersion of 450 \kms~(NGC~1277) and 300 \kms~(IC~843) for display purposes only. The dotted blue line is the global spectrum for each galaxy. Blue and red shaded regions show the index band and continuum definitions respectively. }
\label{fig:spectra}
\end{figure}

\section{Telluric Correction and Sky Subtraction}
\label{sec:sky_subtraction}

At the redshift of these galaxies, telluric absorption is prevalent around the MgI and TiO features in both objects and near the blue continuum band of the NaI feature in NGC~1277. We used the ESO tool  {\sc molecfit}~\citep{Molecfit} to remove it from our spectra. {\sc molecfit} creates a synthetic telluric absorption spectrum based on a science observation contaminated by telluric absorption. Using the radiative transfer code of \citet{Clough2005}, a model line-spread function of the instrument used to observe the data and a model atmospheric profile based on the temperature and atmospheric chemical composition at the time and place of observation, a telluric spectrum is fit to the science spectra and then divided out. We use {\sc molecfit} between the regions $\lambda \lambda$ 7561-7768 \AA, 81212-8338\AA ~and 8931-9875 \AA. 
 
Variations in night sky emission lines occur on similar timescales to our observations, meaning that significant residuals from telluric emission remain after first order sky subtraction. This is especially true in the far red end of the spectrum. These residuals are the main source of systematic uncertainty in the measurement of the FeH band, and so must be accurately subtracted to ensure robust index measurements at 1$\mu m$. We use two independent sky subtraction methods in this work: removing skylines whilst simultaneously fitting kinematics, and fitting each wavelength slice of our observation cubes with a model galaxy profile and sky image before subtracting the best fit sky model.

\subsection{Removing Skylines with \texttt{pPXF}}
\label{sec:pPXF_Sky_Sub}

The first sky subtraction technique uses the method of penalised pixel fitting \citep[\texttt{pPXF}]{ppxf} to fit sky spectra to our data at the same time as fitting the stellar kinematics, as discussed in \citet{2009MNRAS.398..561W} and \cite{2017MNRAS.465..192Z}. This involves passing \texttt{pPXF} a selection of sky templates (as well as stellar templates) which are scaled to find the best fit linear combination to the remaining sky residuals. 

The sky templates were extracted from the dedicated sky frames observed throughout the night. To account for instrument flexure, each sky template was shifted forward and backwards in wavelength by up to 2.5 pixels (2.5 \AA).  Note that the \texttt{pPXF} sky subtraction occurs \textit{after} first order sky-subtraction, and so we also include negatively-scaled sky spectra in the list of templates in order to fit negative residuals (which correspond to over-subtracted skylines). The sky spectra were also split into separate regions around emission lines caused by different molecular transitions, based on definitions from \cite{2007MNRAS.375.1099D}. We also introduced a small number of further splits to the sky spectrum by eye, around areas where skyline residuals changed sign. Each region was allowed to vary individually in \ppxf~to achieve the best sky subtraction. 

The choice of sky splits makes a noticeable difference to the quality of sky subtraction, especially around the feature most contaminated by sky emission, the Wing-Ford band. Correspondingly, the sky split selection has a non-negligible effect on the FeH index measurement. We selected the total number and location of cuts to the sky spectrum around FeH by quantifying the residuals of the sky subtracted spectrum around the best-fitting \ppxf~template, for various sky split combinations. We chose the combination of sky splits which had a distribution containing fewest catastrophic outliers (i.e most similar to a normal distribution), both by eye and quantified using the Anderson-Darling test statistic \citep{AndersonDarling}. This process is discussed in further detail in Appendix \ref{Appendix:SkySub}.

\subsection{Median Profile Fitting}

The second sky subtraction method is independent of the first. Each observation cube (which has undergone first order sky subtraction) is a combination of galaxy light and residual sky light. In each wavelength slice, sky emission corresponds to an addition of flux in all pixels whereas galaxy light is concentrated around the centre of the observation. We aim to model these two contributions in a single data cube and subtract off the best fitting sky model.

We take the median image of the data cube as the galaxy model in our fitting procedure. This assumes that the shape of galaxy light profile doesn't change over the SWIFT wavelength range of 6300 \AA~to 10412 \AA, but is only scaled up and down as the galaxy gets brighter or dimmer and the instrument throughput varies. The sky model is a flat image at every wavelength slice; the same constant value across the IFU in each spatial dimension. 

Using a simple least squares algorithm, we simultaneously fit the galaxy and sky model to each wavelength slice of an individual cube. We then subtract the best fit sky residuals for each cube, combine the observation cubes together and are left with an alternative sky subtracted data cube for each galaxy. These are binned and passed to \texttt{pPXF} to measure the kinematics as before, except without using the sky subtraction technique of Section \ref{sec:pPXF_Sky_Sub}.

The median profile fitting method leads to slightly noisier results than using \ppxf, and as such all index measurements quoted in this paper are derived from the first sky subtraction method. However our conclusions are unchanged regardless of which sky subtraction technique we employ.  A comparison of the two methods is presented in Appendix \ref{Appendix:SkySub}.

\section{Index Measurements}
\label{sec:index_corr_factor}

To attain a signal to noise (SN) ratio high enough to robustly measure equivalent widths, we binned the data cubes into elliptical annuli of uniform SN, which were then split in half along the axis of the galaxy's rotation. The kinematics in each bin were measured using \ppxf, after which each half of the same annulus was interpolated back to its rest frame and added together. This leads to a roughly constant SN in each bin for each index. Spectra of the FeH and NaI IMF sensitive indices studied in this work, for each radial bin in both galaxies, are shown in Figure \ref{fig:spectra}. 

\begin{figure}
\centering
\includegraphics[width=\linewidth]{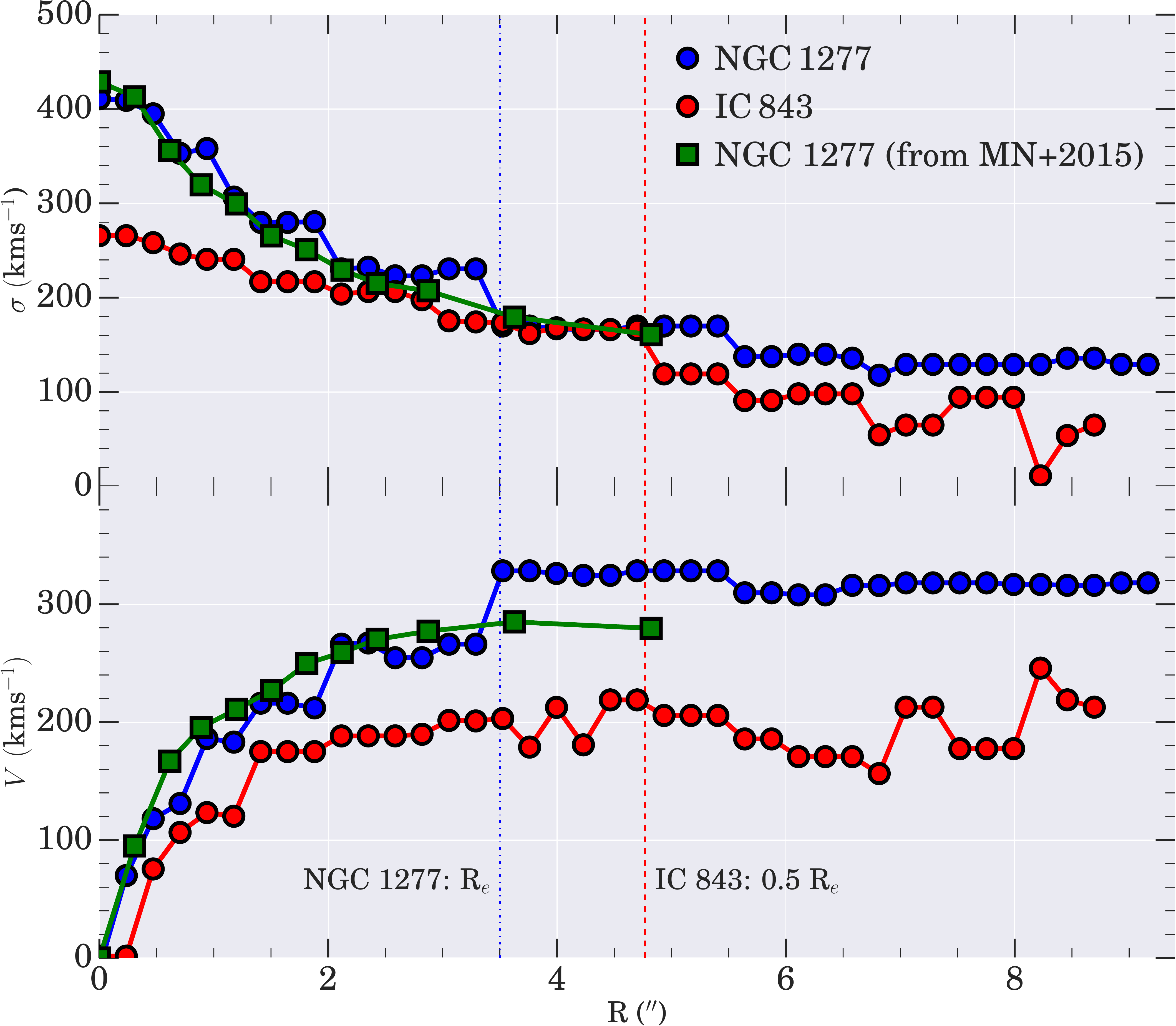}
\caption{Velocity and velocity dispersion parameters for IC~843 (red) and NGC~1277 (blue). Both galaxies show large central velocity dispersions (especially NGC~1277, with $\sigma_0$=410\kms) and ordered rotation at larger radii. Green points are long-slit observations of NGC~1277 taken from MN15.}
\label{fig:1DKinematics}
\end{figure}

We also make velocity and velocity dispersion measurements as a function of radius by binning the datacube to a SN ratio of 15 (for NGC~1277) or 20 (for IC~843), then place a pseudo-slit across the cube along the major axis of each galaxy. These are shown in Figure \ref{fig:1DKinematics}, along with the long-slit results from MN15. Both galaxies are fast rotators, with peak rotation velocities reaching $\pm$ 300 \kms~in NGC~1277 and $\pm$ 200 \kms~in IC~843. The central velocity dispersion in NGC~1277 is remarkably high at $\sim$420\kms, in agreement with the values measured by MN15. 

The equivalent widths of absorption features depend on the velocity dispersion of the spectrum they are measured from. A larger velocity dispersion tends to "wash out" a strong feature, leading to a smaller equivalent width. In order to compare measurements between different radii in the same galaxy, as well as between separate galaxies, we correct each index measurement to a common $\sigma$ of 200 \kms~using the same method as \cite{2017MNRAS.465..192Z}.

Equivalent widths are measured using the formalism of \citet{Cenarro2001}, which measures indices relative to a first order error-weighted least squares fit to the pseudo-continuum in each continuum band. We propagate errors from the variance frames of each observation by making a variance spectrum for each science spectrum. All error bars in this work show 1$\sigma$ uncertainties.

\label{sec:IndexCorrectionFactors}

\subsection{Selected spectral features}

The SWIFT wavelength range extends from 6300\AA~to 10412\AA, covering the IMF sensitive indices NaI0.82, CaT0.86 and FeH0.99. We also make radial measurements of the TiO0.89 bandhead and the MgI0.88 absorption feature. 

The sodium feature at 0.82 $\mu$m is well studied, with a long history of measurements in the context of IMF measurements \citep[e.g][]{1971ApJS...22..445S, 1980LicOB.823....1F, 1997ApJ...479..902S}. It is strengthened in the spectra of dwarf stars and is sensitive to the abundance of sodium \citep{CvD12a}. The feature is a doublet in the spectra of individual stars, but the velocity dispersion in  massive galaxies often blends it into a single feature.

The Wing-Ford band feature is a small absorption feature of the Iron Hydride molecule at 0.99$\mu$m \citep{1969PASP...81..527W}. It is particularly sensitive to the lowest mass dwarf stars, weakens in [Na/H] enhanced populations and is relatively insensitive to $\alpha$-abundance \citep{CvD12a}.

The Calcium Triplet is the strongest absorption feature studied in this work, and is IMF sensitive due to the fact that it is strong in giant stars but weak in dwarfs. Its use as an IMF sensitive index was studied in \cite{2003MNRAS.339L..12C}, where an anti-correlation between the CaT equivalent width and $\log(\sigma_0)$ was presented. Calcium is also an $\alpha$ element, although interestingly the Ca abundance has been shown to be depressed with respect to other $\alpha$ elements by up to factors of two in massive ETGs \citep{2003MNRAS.343..279T}. The feature also weakens in spectra with enhanced [Na/H], and is sensitive to the [Ca/H] abundance ratio. 

The TiO0.89 and MgI0.88 features are both relatively insensitive to the IMF. In the models of \citet{CvD12a}, the TiO bandhead is strongly sensitive to the $\alpha$-enhancement of the population, as well as weakening with older stellar ages. It also becomes stronger with increased [Ti/Fe] and weaker with [C/Fe]. The MgI0.88 feature displays the opposite behaviour with respect to stellar age, becoming stronger as a population ages, and becomes deeper with increasing [Mg/Fe] and [$\alpha$/Fe]. 

\section{Results}
\label{sec:results}

\subsection{Radial variation in index strengths}

\label{sec:IndexResults}
\begin{figure*}
\centering

\includegraphics[width=0.81\linewidth]{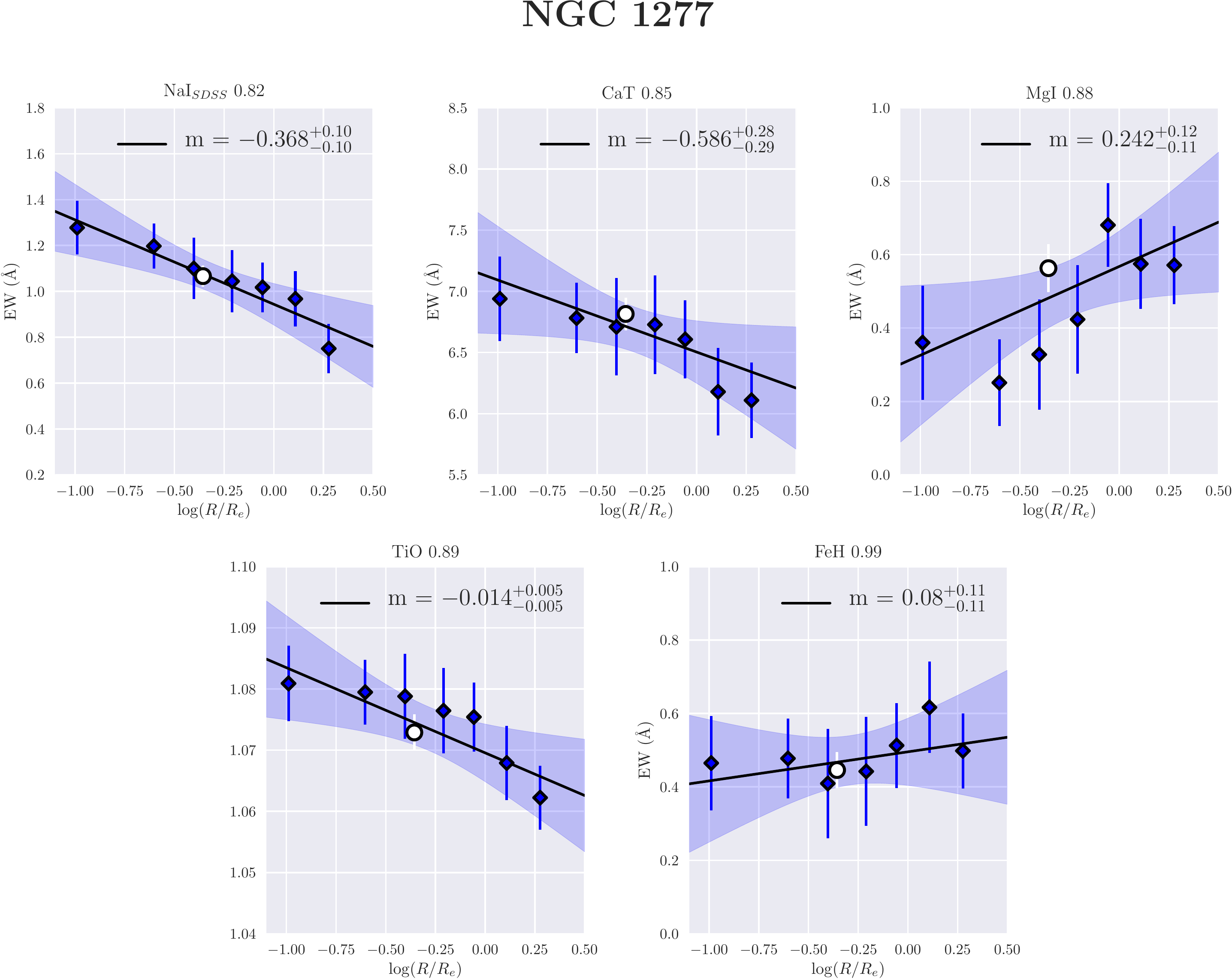}
\includegraphics[width=0.81\linewidth]{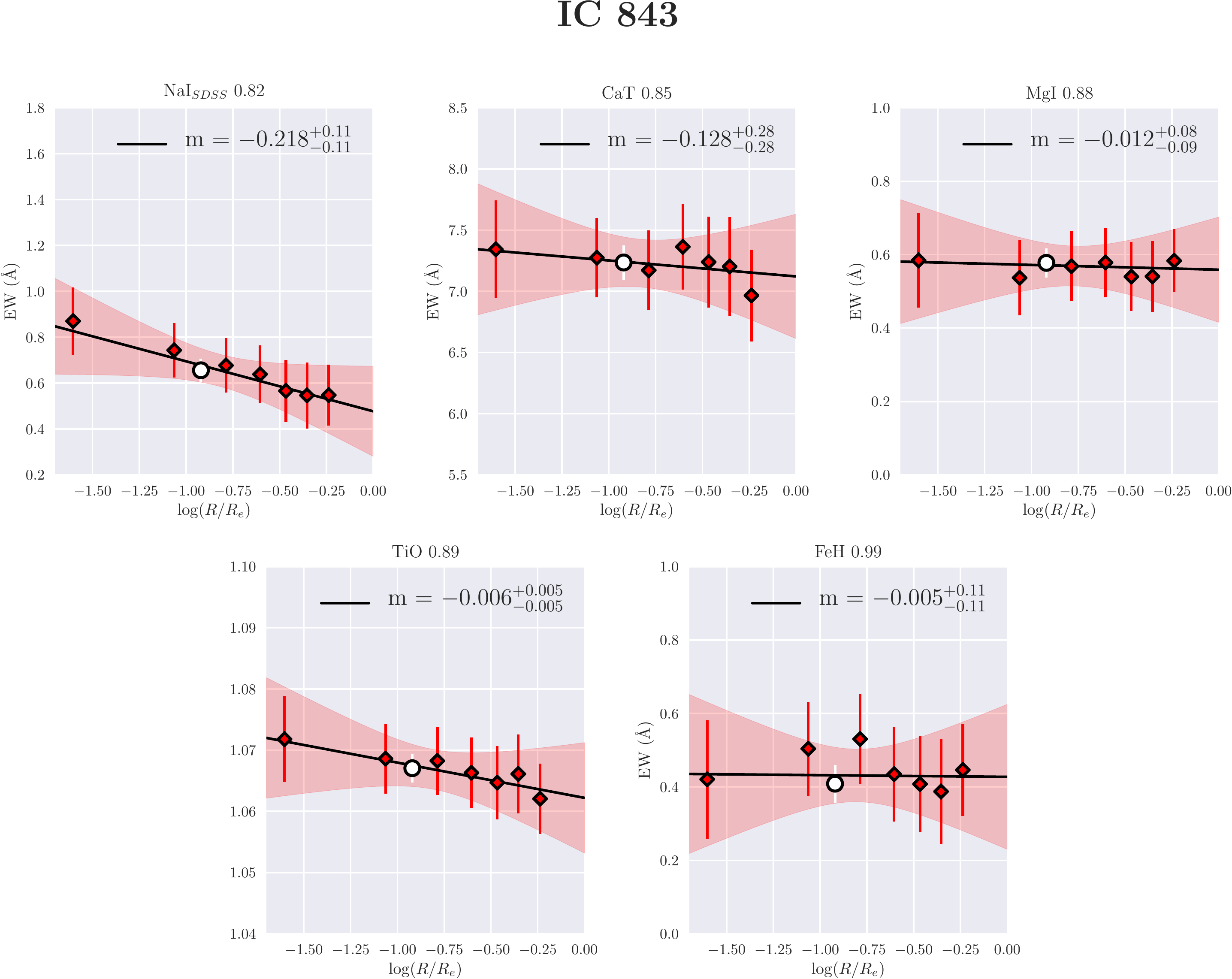}
\caption{Radial measurements for each index in NGC~1277 (blue, top) and IC~843 (red, bottom). All index measurements have been corrected to a common velocity dispersion of 200\kms~(see section \ref{sec:index_corr_factor}). $1\sigma$ errors around the best fit line (encompassing uncertainty in both the gradient and intercept) are shown as shaded regions. The value of the best-fit gradient, $m$, is shown in each panel. The white circle shows the value of each index in the integrated spectrum of each galaxy. The individual radial index measurements for each galaxy are presented in Tables \protect\ref{tab:all_inds_NGC1277} and \protect\ref{tab:all_inds_IC843}}
\label{fig:AllIndices}
\end{figure*}

Figure \ref{fig:AllIndices} shows the results of measuring the IMF sensitive absorption features in NGC~1277 and IC~843 as a function of radius. As discussed in Section \ref{sec:IndexCorrectionFactors}, these measurements were taken at the intrinsic velocity dispersion of the radial bin and then corrected to 200 \kms~for both galaxies. All results are equivalent widths, in units of \AA~and found using the formalism of \citet{Cenarro2001}, except for that of TiO which is a ratio of the blue and red pseudo-continuua.  Table \ref{tab:Index_Gradients} gives the best fitting gradient, $m$, of the straight line fit to each index, with $1\sigma$ uncertainties from the marginal posterior of $m$. It also lists the measured values of each index in the integrated spectrum of each galaxy. The individual radial index measurements in both galaxies are presented in Tables \ref{tab:all_inds_NGC1277} and \ref{tab:all_inds_IC843}.

\begin{table}
	\centering
	\caption{Measured index trends with respect to $\log(R/R_{e})$, with $1\sigma$ uncertainties. Gradient units are \AA / $\log(R/R_{e})$, apart from the TiO index gradient which is simply 1/$\log(R/R_{e})$. Index measurements are in \AA, apart from the TiO index which is a ratio of pseudo-continuua  }
	\label{tab:Index_Gradients}
	\begin{tabular}{lcccc}
    		\hline
		Index&\multicolumn{2}{c}{Best fit gradient} & \multicolumn{2}{c}{Global Index Value} \\
		& IC~843& NGC~1277 & IC~843& NGC~1277\\
		\hline

NaI& $-0.218^{+0.11}_{-0.11}$&  $-0.368^{+0.10}_{-0.10}$ & $0.66\pm0.05$&$1.07\pm0.05$ \\
CaT & $-0.126^{+0.28}_{-0.28}$& $-0.595^{+0.29}_{-0.30}$& $7.24\pm0.14$ &  $6.81\pm0.13$    \\
MgI & $-0.012^{+0.09}_{-0.09}$& $0.231^{+0.12}_{-0.12}$& $0.58\pm0.04$ & $0.56\pm0.06$  \\
TiO & $-0.006^{+0.005}_{-0.005}$&  $-0.011^{+0.005}_{-0.005}$&  $1.067\pm0.002$ & $1.073\pm0.002$  \\
FeH & $-0.015^{+0.11}_{-0.11}$&  $0.081^{+0.11}_{-0.10}$& $0.41\pm0.05$  & $0.44\pm0.05$ \\

	\end{tabular}
\end{table}

The most significant index gradient in NGC~1277 is in NaI, which drops from 1.3 \AA~in the very centre to $\sim 0.8$ \AA~at 1.9R$_{e}$. This behaviour is consistent with the findings of MN15, who found a similarly strong radial gradient in this object. We also measure negative gradients in the CaT (at a $\sim2\sigma$ significance) and the TiO0.89 ($2.75\sigma$). The measurements of MgI0.88 index show a positive trend with radius, although the scatter in these measurements is large, possibly due to the effects of residual telluric absorption. We do not find evidence for a radial gradient in FeH0.99 in NGC~1277, with the gradient in index strength being consistent with zero. 

The most significant gradient in IC~843 is also the NaI feature, albeit offset to a weaker index strength. We also see a significant radial trend in TiO0.89. We measure flat radial profiles for MgI0.88, the CaT and the Wing Ford band, with all three indices having a best fit gradient fully consistent with zero.

\section{Analysis}
\label{sec:Analysis}

\subsection{Stellar population synthesis models}
\label{sec:SPS_models}
\begin{figure*}
\centering
\includegraphics[width=0.91\linewidth]{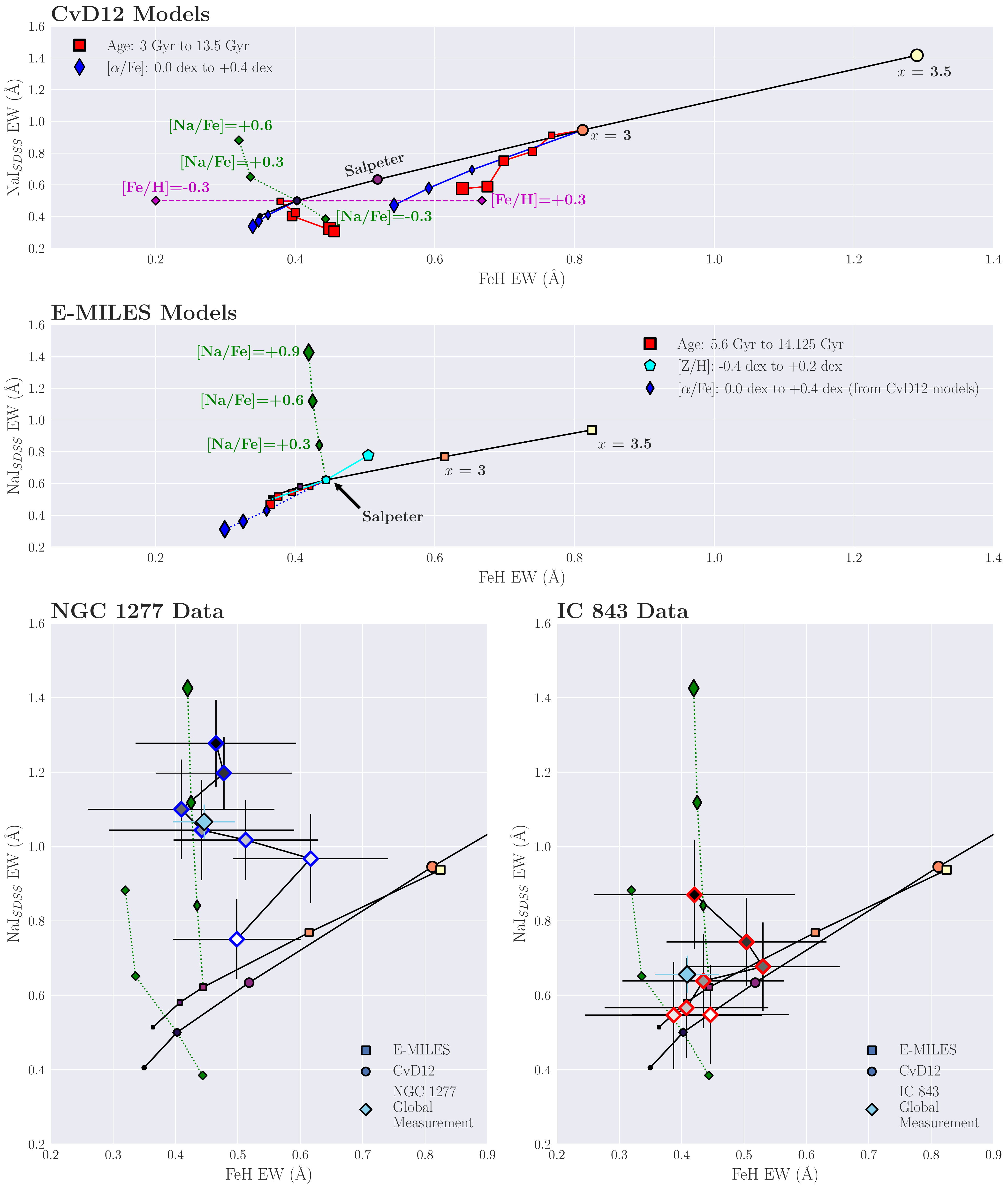}

\caption{A comparison of our measurements of NaI and FeH equivalent widths to two sets of stellar population models. All equivalent widths are measured at a common velocity dispersion of 200\kms. The upper two panels show the change in FeH and NaI equivalent width with varying stellar population parameters in the CvD12 models (first panel) and E-MILES models (second panel). The black line in both panels shows index responses to varying the IMF slope, $x$, in an old stellar population at solar metallicity, [$\alpha$/Fe]=0.0 and with solar elemental abundance ratios. Note the differences in index equivalent width predictions, for the same IMF slope, between the models. Changes in [$\alpha$/Fe] from solar to +0.4 dex (small to large blue diamonds), and age from 3 to 13.5 Gyr (large to small red squares) from a Chabrier and $x=3$ IMF are shown in the top panel, for the CvD12 models. Also plotted are predictions for abundance variations of +0.6 dex to -0.3 dex in [Na/Fe]  (green) and [Fe/H] (purple) from a Chabrier IMF. The second panel shows variations of the E-MILES model predictions for ages from 5.6 to 14.125 Gyr (large to small red squares), [Z/H] from -0.4 to +0.2 dex (small to large cyan pentagons) and [Na/Fe] from solar to +0.9 dex (small to large green diamonds), all from a Salpeter IMF.  We also include response functions from CvD12 showing variations in [$\alpha$/Fe] abundance. 
The lower two panels show our resolved measurements in NGC~1277 (left, blue outline) and IC~843 (right, red outline), coloured by their radial position from the centre of the galaxy (dark) to the outskirts (light). The global measurements for both galaxies are plotted in light blue. Index predictions from the models vary from a bottom light IMF (dark colours, bottom left) to a bottom heavy IMF (light colours, top right), with equivalent IMF slopes plotted in the same colour; for example, the predictions for an IMF slope of $x=3$ are coloured orange for both models. See Section \ref{sec:SPS_models} for discussion.}
\label{fig:sps_models}
\end{figure*}

Figure \ref{fig:sps_models} shows a comparison of our NaI and FeH measurements to two sets of stellar population synthesis (SPS) models, each convolved to 200 \kms~to match our measurements. The top two panels show index predictions from the CvD12 \citep{CvD12a} and E-MILES \citep{1996ApJS..106..307V} libraries for changes in IMF slope from an old (13.5 Gyr for the CvD12 models, 14.125 Gyr for E-MILES) stellar population at solar metallicity, $\alpha$-abundance and elemental abundance ratios.  Also shown are variations in index strength with a variety of stellar population parameters included in each set of models. The bottom two panels show our measurements of NGC~1277 (left) and IC~843 (right), along with CvD12 and E-MILES model predictions. 

The CvD12 models allow changes in [$\alpha$/Fe], age and the abundance ratios of various elements, with [Na/Fe] and [Fe/H] being most important to us here. A change in [Fe/H] of $\pm$0.3 dex has no effect on the predicted NaI equivalent width, whilst understandably leading to a large variation in FeH strength. The result of increasing [Na/Fe] is a strengthening of the NaI index combined with a weakening of the FeH equivalent width. This is due to the fact that Na is an important electron contributor in cool giant and dwarf stars, and large abundances of Na in these stellar atmospheres tends to encourage the dissociation of molecules like FeH. An $\alpha$-enhanced population leads to weaker FeH and NaI predictions, especially at steeper IMF slopes, whilst the response of the FeH index to changes in population age is found to be a function of the IMF slope. A full discussion of these SPS models can be found in \cite{CvD12a}.

The E-MILES models  include changes in age and metallicity. A metallicity of +0.2 dex above solar leads to increased equivalent widths for both NaI and FeH, whilst younger ages tend to weaken both indices. \cite{2017MNRAS.464.3597L} have produced the "Na-MILES" models, which are SPS templates spanning the E-MILES wavelength range with enhanced [Na/Fe] abundance ratios of up to 1.2 dex.  Interestingly, these templates predict that the FeH index is less sensitive to the effect of [Na/Fe] enhancement than the CvD12 models. We also expand the dimensionality of these models by applying response functions  for changes in [$\alpha$/Fe] from the CvD12 models, in a similar way to \cite{2015ApJ...803...87S}.

Figure \ref{fig:sps_models} also highlights a complicating factor in our interpretation of our NaI and FeH measurements: the different index predictions from the CvD12 and E-MILES models for the same value of IMF slope. The largest difference is for the most bottom heavy IMFs: the $x=3.5$ IMF slope prediction for FeH is $\sim37\%$ weaker in the E-MILES models compared to the CvD12, whilst the NaI predicted equivalent width is $\sim39\%$ smaller. A large part of this difference is due to the different low-mass cutoff, $m_{c}$, assumed for the IMF in each case: 0.08 $M_{\odot}$ in the CvD12 models and 0.1 $M_{\odot}$ for E-MILES. The CvD12 models therefore have a larger number of very low-mass stars and predict stronger NaI and FeH equivalent widths. 

Constraining the low-mass cut off of the IMF is a technically demanding task, with recent measurements of $m_{c}$ by \cite{2015MNRAS.452L..21S} and \cite{2013MNRAS.436..253B} combining modelling of IMF sensitive indices with constraints from strong gravitational lensing and dynamics. Since in this work we are unable to distinguish between $m_{c}=0.08$ $M_{\odot}$ and $m_{c}=0.1$ $M_{\odot}$, we conduct our analysis and draw conclusions using both sets of SPS models. 

 Note that there are other key differences between the model spectra, largely due to the different ways they are computed. The two sets of models use different isochrones for the lowest mass stars, as well as different methods to attach stars to these isochrones. Figure \ref{fig:sps_models} shows that the response of both NaI and FeH to increases in [Na/Fe] abundance is  different for the two sets of models, and at fixed [Z/H] these indices are also more sensitive to changes in [$\alpha$/Fe] in the E-MILES models than CvD12. A comprehensive comparison of the IMF-sensitive features below 1$\mu m$ in the two sets of models is presented in \cite{2015ApJ...803...87S}.

The measurements in both galaxies scatter around similar areas of parameter space: above a Salpeter IMF in the direction of [Na/Fe] enhancement, with the NGC~1277 points further from the model lines than IC~843. Notably, our measurements disfavour very bottom heavy single power law IMFs with $x$>3 in both NGC~1277 and IC~843.

\subsection{IMF determination from global measurements}
\label{sec:IMF_values}

\subsubsection{Spectral Fitting}
\label{sec:spectral_fitting}

\begin{figure*}
\centering
\includegraphics[width=\linewidth]{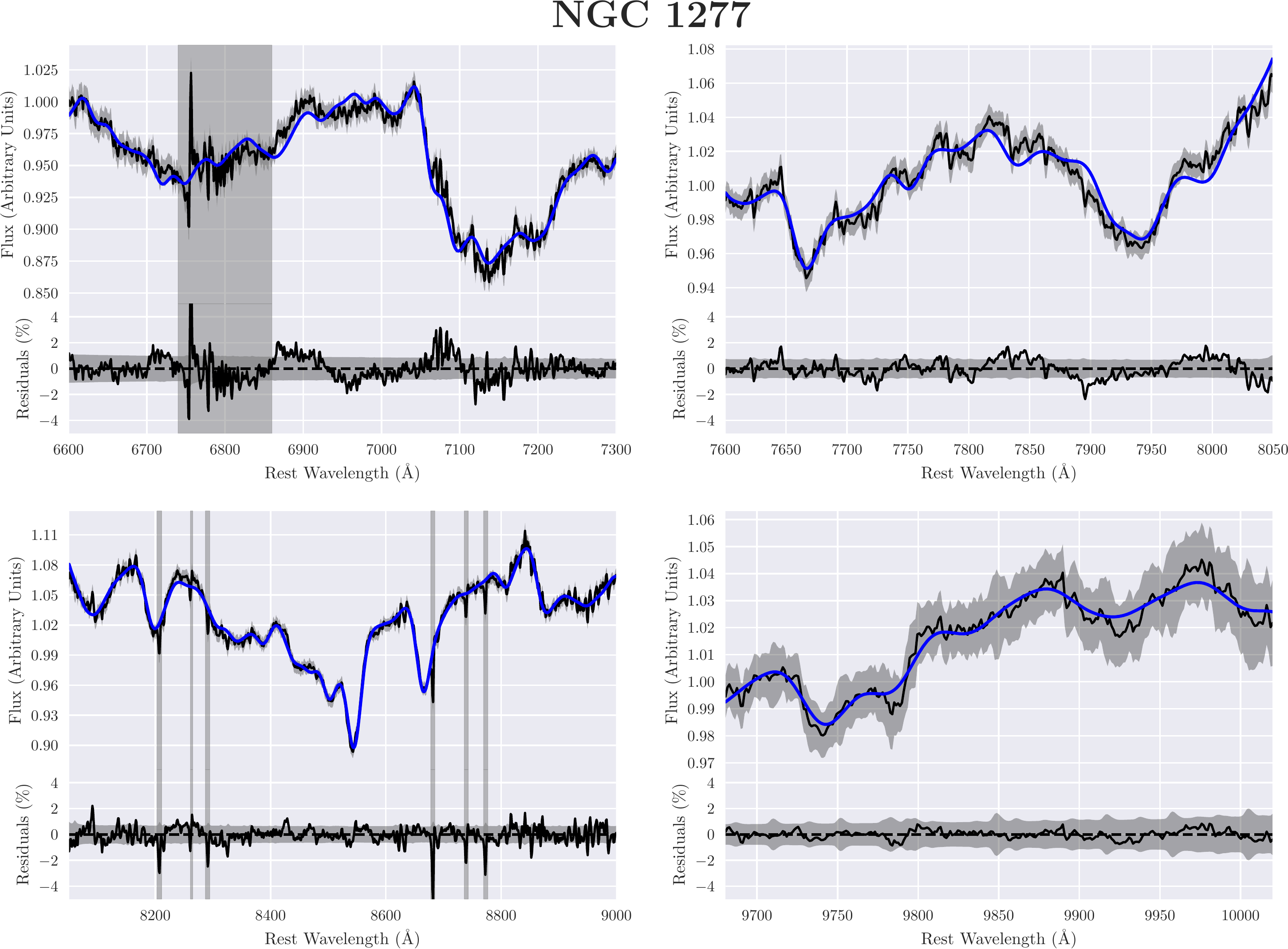}
\caption{Fit to the global spectrum of NGC~1277.  Within each panel, the upper plot show the data and best-fitting template whilst the lower plot shows the residuals between the data and the fit. Grey shaded regions show the noise level of the data.  }
\label{fig:SpectralFit_NGC1277}
\end{figure*}

\begin{figure*}
\centering
\includegraphics[width=\linewidth]{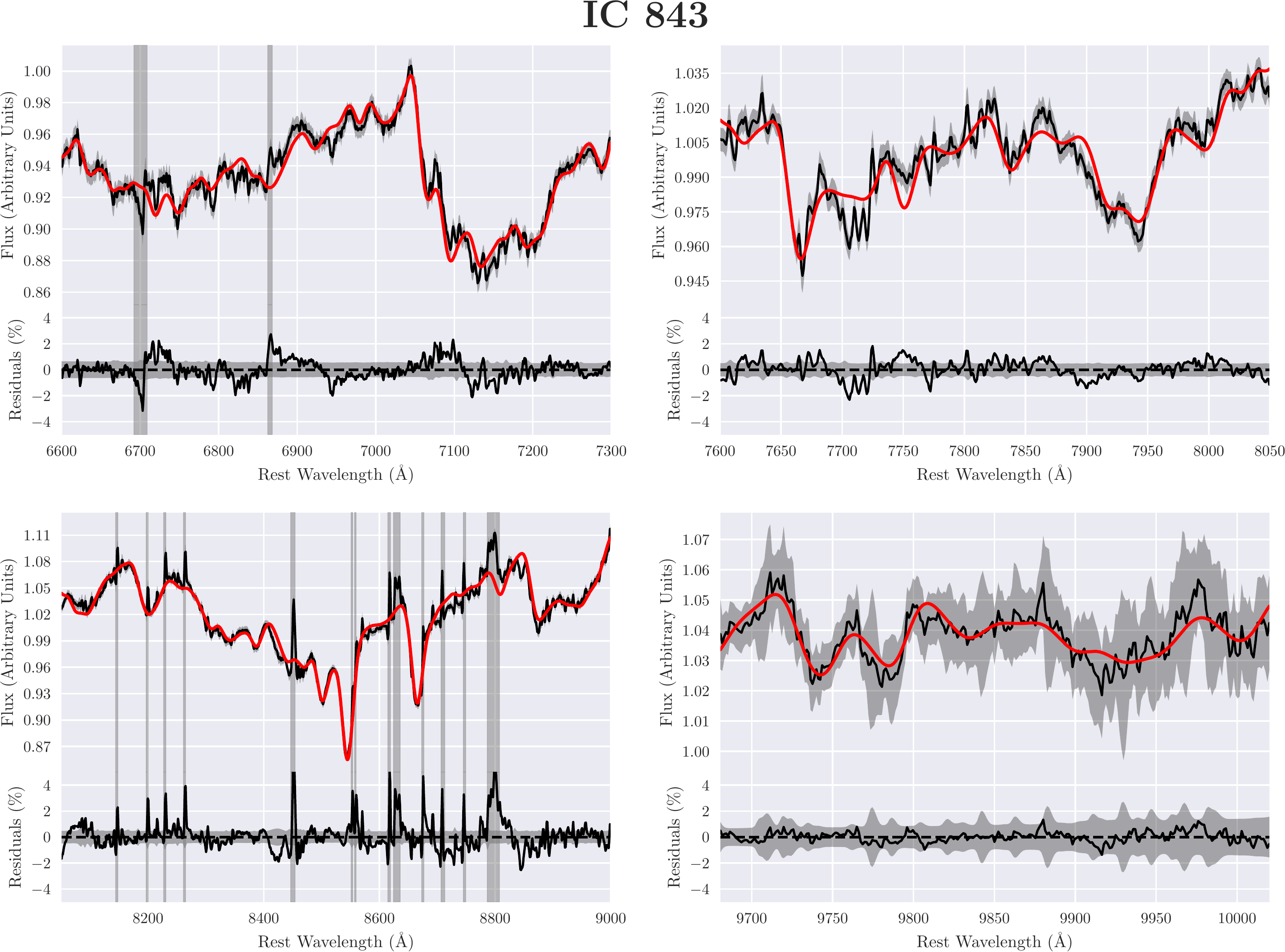}
\caption{Fit to the global spectrum of IC~843: see the caption of Figure \ref{fig:SpectralFit_NGC1277} for details.}
\label{fig:SpectralFit_IC843}
\end{figure*}

In order to make quantitative statements about the global IMF slope in these objects, we fit templates from the spectral library from CvD12 to the integrated spectrum from each galaxy. This technique is discussed extensively in CvD12, and our approach is very similar. The spectral fitting covers wavelengths from 6600 \AA~to 10020 \AA~in the rest frame of each galaxy, split into four sections: 6600-7300 \AA, 7600-8050 \AA, 8050-9000 \AA~and 9680-10020 \AA. The two gaps between 7300-7600 \AA~and 9000-9680 \AA~were chosen to avoid areas of residual telluric absorption. We have also carefully masked pixels contaminated by sky subtraction residuals in each spectrum. These masked regions are shaded in Figure \ref{fig:SpectralFit_NGC1277} and \ref{fig:SpectralFit_IC843}.

To correct for different continuum shapes between the templates and the data, we fit Legendre polynomials to the ratio of the template and the galaxy. The order of these polynomials is defined to be the nearest integer to $(\lambda_{\mathrm{upper}}-\lambda_{\mathrm{lower}})/100$. We have ensured that slightly varying the order of this polynomial has negligible effect on our results, and have included the effect of this variation in the error budget for each parameter. 

We allow eight parameters to vary when performing the fit: the redshift and velocity dispersion of the template; the chemical abundances [Na/Fe], [Fe/H], [Ca/Fe], [Ti/Fe] and [O/Fe]; and the unimodal IMF slope. Ideally, the stellar age and [$\alpha$/Fe] abundance would also be included as free parameters of the fit. We did not find these quantities to be well constrained by the data, however, since our wavelength coverage does not include many of the blue absorption indices sensitive to these parameters. To overcome this, we use previously published measurements of blue "Lick" indices (which are, to first order, insensitive to the IMF) from \cite{2017MNRAS.tmp..177F} for NGC~1277 and \cite{Price2011} for IC~843 and the SPS models of  \cite{Thomas2011} to infer stellar age and [$\alpha$/Fe] abundances in these objects. We then fix the values of these parameters during the fit. The derived values are shown in Table \ref{tab:Spectral_Fitting_results} for both galaxies. 

Note that the measurements from \cite{2017MNRAS.tmp..177F} are spatially resolved, and we have matched these data to the aperture size used in this work. Measurements from \cite{Price2011} come from an SDSS fibre covering a diameter of 3\arcsec~in the very centre of IC~843, however, smaller than the 12\arcsec~which contribute to our global spectrum. Such a discrepancy is unavoidable, but does mean that if any radial gradients in the stellar age or [$\alpha$/Fe] abundance exist in IC~843 then the parameters assumed for the global spectrum would be incorrect. Furthermore, the derived age and [$\alpha/Fe$] for NGC~1277 and IC~843 were found using a different set of stellar population models than were used for the spectral fitting (the \cite{Thomas2011} models rather than those from CvD12).
Small systematic offsets between the models could exist, implying that our fixed values found from the \cite{Thomas2011} models may not be appropriate for the fitting using the CvD12 templates.

To account for this, we have also computed fits (for both galaxies) where we varied these assumed values of age and [$\alpha$/Fe] abundance. The resulting change in the derived parameters are included in the error budget for each result (see Table \ref{tab:Spectral_Fitting_results}).

The CvD12 models also allow for variation in further elemental abundances, as well as an "effective isochrone temperature" nuisance parameter, $T_{\mathrm{eff}}$.  $T_{\mathrm{eff}}$ slightly changes the isochrone each galaxy template is built from, which is a proxy for varying metallicity (since variations in total [Z/H] are not modelled in the CvD12 library). Further discussion can be found in CvD12. We compute fits (at each fixed age and [$\alpha$/Fe] abundance assumed above) which include all further element variations (in [C/Fe], [N/Fe], [Mg/Fe] and [Si/Fe]), as well as $T_{\mathrm{eff}}$, to ensure that the low-mass IMF slope we recover is not being driven by an elemental abundance variation we have neglected. We find that the best-fit IMF is negligibly affected in either galaxy. Any variations in the derived parameters as a result of this process are also folded into the uncertainties reported in Table \ref{tab:Spectral_Fitting_results}.

The fit was performed using the Markov-Chain Monte-Carlo ensemble-sampler \texttt{emcee} \citep{2013PASP..125..306F}. We use a simple, $\chi^{2}$ log-likelihood function with flat priors on each parameter. 400 "walkers" explore the posterior probability distribution, each taking 10,000 steps, giving $4\times10^{6}$ samples in total. We discard the first 8000 steps of each walker as the "burn-in" period. Each chain was inspected for convergence and we have run tests to ensure that chains which start from different areas of parameter space converge to the same result. 

Results of the fitting are shown in Figures \ref{fig:SpectralFit_NGC1277} and \ref{fig:SpectralFit_IC843}, with prior ranges and derived quantities shown in Table \ref{tab:Spectral_Fitting_results}. The fits to the spectra are good, with residuals at around the $1\%$ level for the majority of the wavelength range. The reduced $\chi^{2}$ values are 0.41 and 0.95 for NGC~1277 and IC~843 respectively. We note, however, that the spectral range from 7600 to 8050 \AA~in IC~843 shows significant residuals. We have ensured that our conclusions for both galaxies are unaffected if we remove this region from the fit, and included the small variations in derived parameters in our error budget.

We find both galaxies to have super-solar [Na/Fe] abundances by factors of between 3 and 5, with NGC~1277 requiring greater Na enhancement than IC~843. NGC~1277 also has a more bottom heavy low-mass IMF slope than IC~843, with best fitting single power law IMF slopes of  $2.69 ^{+0.11}_{-0.11}$ for NGC~1277 and $2.27^{+0.16}_{-0.18}$ for IC~843.

The magnitude of the [Na/Fe] enhancement in both objects is large, but it should also be noted that the spectral response to increases in [Na/Fe] between the CvD12 and E-MILES models is markedly different (as shown in Figure \ref{fig:sps_models}). This implies that the magnitude of the super-solar [Na/Fe] abundance is likely to be model dependent.

\begin{table}
	\centering
	\caption{Stellar population parameters for NGC~1277 and IC~843, derived from spectral fitting. The stellar age and [$\alpha/$Fe] abundance (italicised) were derived from optical index measurements from \protect\cite{2017MNRAS.tmp..177F} (for NGC~1277) and \protect\cite{Price2011} (for IC~843) and kept fixed during the fit. Errors are a combination of photon errors,  marginalisation over changes in the fixed parameters, inclusion of further element variations, small changes in the multiplicative polynomial order and the removal of the 7600-8050 region of the spectrum.  (see Section \ref{sec:spectral_fitting}).}
	\label{tab:Spectral_Fitting_results}
	\begin{tabular}{lccc}
    		\hline
		Parameter & NGC~1277 & IC~843&Prior\\
		\hline
\textit{Age (Gyr)} & $13^{+1}_{-3}$ & $10\pm3$&---\\ \relax
\textit{[$\alpha$/Fe]}& $0.3\pm0.1$ & $0.3\pm0.1$&--- \\\hline\relax
$\sigma$ (\kms)& $377^{+8}_{-8}$ &  $287^{+8}_{-7}$& [0, 1000] \\\relax
[Na/Fe] & $0.71 ^{+0.30}_{-0.26}$ & $0.49^{+0.17}_{-0.17}$& [-0.3, 0.9]\\\relax
[Fe/H] & $0.02^{+0.07}_{-0.07}$ & $-0.10^{+0.06}_{-0.06}$& [-0.3, 0.3]\\\relax
[Ca/Fe] & $-0.18^{+0.05}_{-0.04}$ & $-0.20^{+0.02}_{-0.02}$& [-0.3, 0.3]\\\relax
[Ti/Fe] & $-0.22^{+0.18}_{-0.17}$ & $-0.02^{+0.23}_{-0.22}$ &  [-0.3, 0.3]\\\relax
[O/Fe] & $ 0.001^{+0.002}_{-0.002}$ & $0.00^{+0.01}_{-0.00}$ &  [0.0, 0.3]\\\relax
IMF slope & $2.69 ^{+0.11}_{-0.11}$  & $2.27^{+0.16}_{-0.18}$ & [0.0, 3.5]\\
	\end{tabular}
\end{table}

\subsubsection{Index Fitting}
\label{sec:index_fitting}
In order to directly compare to \cite{2017MNRAS.465..192Z}, we also use the global FeH and NaI index measurements of NGC~1277 and IC~843 to make quantitative statements about the IMF slope in each galaxy. By interpolating the predicted FeH equivalent widths from the CvD12 and E-MILES models and comparing to our global FeH measurement, we measure global IMF slopes in each object. Appendix \ref{Appendix:IMF_Calculations} discusses the precise calculations in detail. We assume the same population parameters as in Section \ref{sec:spectral_fitting}, as well as including the effect of the non-solar abundances found from the spectral fitting for each galaxy (see Table \ref{tab:Spectral_Fitting_results}). 

In contrast to the CvD12 models, the E-MILES models allow variation in the total metallicity, [Z/H]. Using index measurements from the same sources as before (\cite{2017MNRAS.tmp..177F} for NGC~1277,  \cite{Price2011} for IC~843), we assume a total metallicity of +0.16 dex for NGC~1277 and +0.0 dex for IC~843 in our global spectra.

\begin{figure*}
\centering
\includegraphics[width=\linewidth]{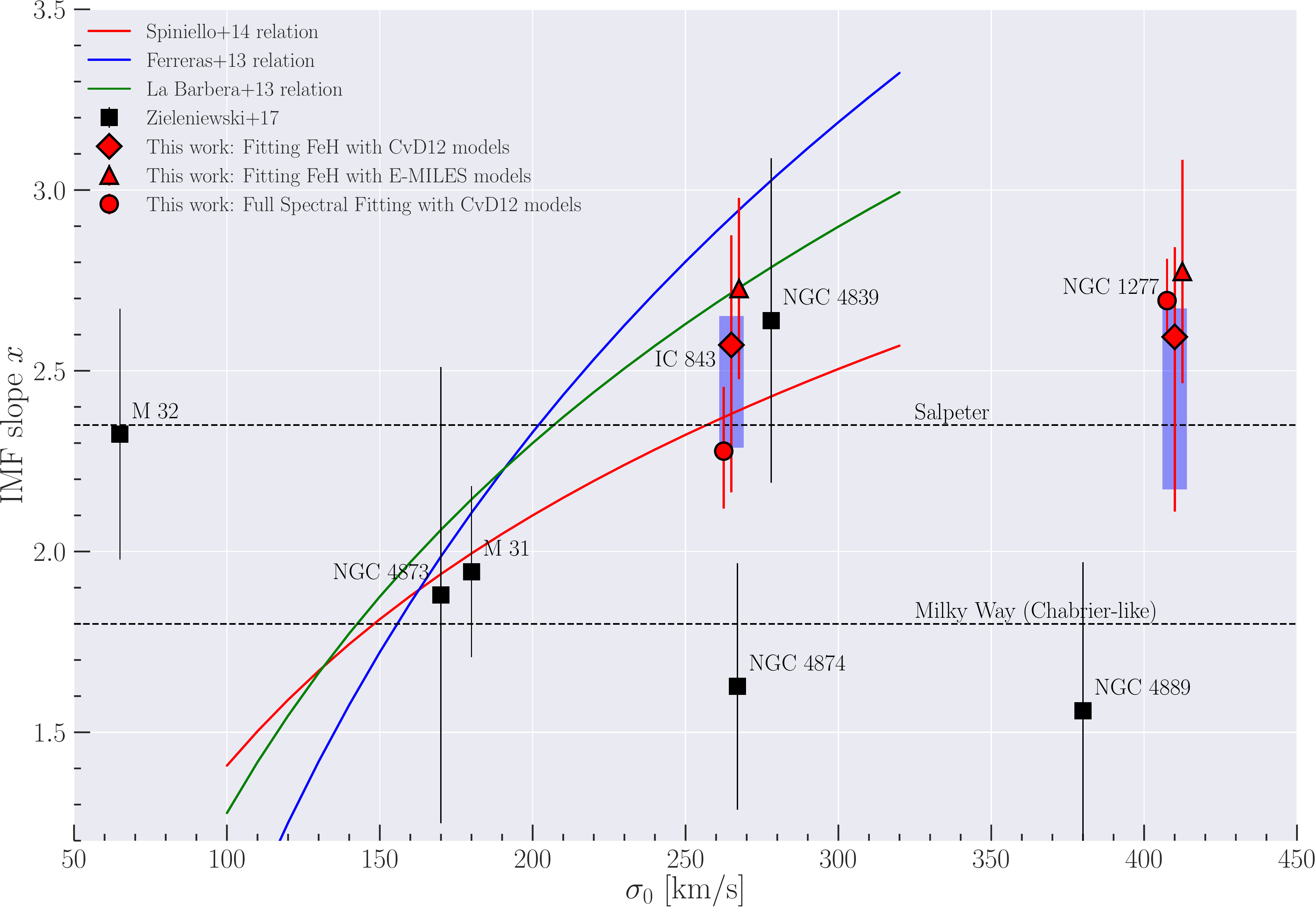}
\caption{Derived IMF slope for both galaxies studied in this work, plotted against their central velocity dispersion. Discussion of the IMF slope calculation is presented in Sections \ref{sec:spectral_fitting} and \ref{sec:index_fitting}, as well as Appendix \ref{Appendix:IMF_Calculations}. Circles denote the values determined from spectral fitting with the CvD12 models. Diamonds correspond to the values derived by fitting the FeH index with the CvD12 stellar population models and triangles denote the values found using the E-MILES models (which are slightly offset in the $x$ direction), following \protect\cite{2017MNRAS.465..192Z}.  See section \ref{sec:IMF_values} for details. The blue shaded regions show the effect of uncertainties in the stellar population parameters assumed for each galaxy in the index fitting. They are derived by assigning a normal distribution to each parameter (centred on the appropriate values and of width 0.1 dex for [Z/H], [Fe/H], [Na/H] and [$\alpha$/Fe], 1 Gyr for age), then drawing 1000 samples and calculating the IMF for each one. The blue regions denote the 16th and 84th percentiles of these samples. Red error bars combine measurement errors with stellar population uncertainties. A similarly sized uncertainty in the population parameters also applies to the E-MILES model points, but is not shown for clarity. Coloured lines show proposed unimodal IMF-$\sigma_0$ correlations from \protect\cite{2013MNRAS.429L..15F}, \protect\cite{2013MNRAS.433.3017L} and \protect\cite{2014MNRAS.438.1483S}, whilst black squares are values from \protect\cite{2017MNRAS.465..192Z}. Note that measurements from \protect\cite{2017MNRAS.465..192Z} utilise the CvD12 models only, and that this work implements an improved handling of response functions to correct for non-solar abundances.}
\label{fig:IMF_sigma}
\end{figure*}

To measure the effect of uncertainty in the assumed stellar population parameters for each galaxy, we modelled each parameter as a normal distribution centred on the values described above. The width of these distributions are 0.1 dex for [Z/H] and [$\alpha$/Fe] and 3 Gyr for stellar age. For [Na/Fe] and [Fe/H], we use the values and errors from Table \ref{tab:Spectral_Fitting_results}. We drew 1000 random samples from the distribution of each parameter, then recalculated the IMF slope in each case. The 16th and 84th percentiles of these samples are plotted as the blue shaded regions in Figure \ref{fig:IMF_sigma}. 

Results from this second IMF determination method show a nice agreement between the E-MILES and CvD12 stellar population models. In IC~843, the index fitting results are best fit with single power-law IMF slopes heavier than Salpeter:  $x_{\mathrm{CvD}}=2.57^{+0.30}_{-0.41}$, whilst $x_{\mathrm{E-MILES}}=2.72\pm0.25$. However, the spectral fitting leads to an IMF slope shallower than Salpeter: $x_{\mathrm{SF}}=2.27^{+0.16}_{-0.18}$ (although the results are consistent within the error bars). For NGC~1277, the three methods agree well: $x_{\mathrm{CvD}}=2.59^{+0.25}_{-0.48}$, $x_{\mathrm{E-MILES}}=2.77\pm0.31$ and $x_{\mathrm{SF}}=2.69 ^{+0.11}_{-0.11}$
 
Figure \ref{fig:IMF_sigma} shows the derived single-power law IMF slope for IC~843 and NGC~1277, plotted against their central velocity dispersion. Diamonds show the IMF slopes derived using the CvD12 stellar population models, whilst triangles show those found using the E-MILES models. Results from spectral fitting are shown as circles. Also shown are measurements from \cite{2017MNRAS.465..192Z}, as well as the proposed correlations between unimodal IMF slope and $\sigma_0$ from \cite{2013MNRAS.429L..15F}, \cite{2013MNRAS.433.3017L} and \cite{2014MNRAS.438.1483S}. 

Note that using values of [Na/Fe] and [Fe/H] derived by fitting the CvD12 models as corrections to the E-MILES models is not strictly correct, due to the differences in the way the models are constructed and their differing responses to changes in abundance patterns. As a first approximation, however, we have shown that doing so gives consistent results. The ideal solution would be to conduct full spectral fitting with both sets of stellar population models, and future work will investigate this further. 

\subsection{M/L values}

Using these global IMF measurements, we derive V-band stellar mass-to-light values for these galaxies. For NGC~1277, we find $(M/L)_V=10.7^{+2.0}_{-1.4}$ from the spectral fitting,  $(M/L)_V=9.5^{+3.1}_{-2.0}$ from the CvD12 index fitting, and $(M/L)_V=11.2^{+5.9}_{-3.3}$ for the E-MILES index fitting. For IC~843, we find  $(M/L)_V=5.1^{+0.8}_{-0.6}$ from spectral fitting, $(M/L)_V$=$7.3^{+3.4}_{-1.9}$ using CvD12 and $(M/L)_V$=$9.1^{+ 5.4}_{- 2.9}$ using E-MILES. Combining these measurements, weighted by the inverse of their variances, gives $(M/L)_V=10.4\pm1.51$ in NGC~1277 and $5.9\pm1.72$ in IC~843.

Using adaptive optics spectroscopy, \cite{2016ApJ...817....2W} find $(M/L)_V=9.3\pm1.6$ in the very centre of NGC~1277, whilst seeing limited observations out to $\sim3R_{e}$ by \cite{2015MNRAS.452.1792Y} find $(M/L)_V=6.5\pm1.5$, under the assumption of a constant stellar $(M/L)$ with radius. 

MN15 infer the V-band stellar $(M/L)_V$ ratio in NGC~1277 to be 7.5 at 1.4 $R_{e}$, rising to 11.6 in the centre, from their analysis of IMF-sensitive absorption features and the assumption of a bimodal IMF. Whilst we are unable to make such a resolved $(M/L)_V$ measurement with our data, these values are generally in agreement with our integrated measurement (which extends out to just over 2.2 $R_e$). 

In IC~843, \cite{Thomas} make a dynamical measurement of the $(M/L)$ in the $R_c$ band, with observations extending to further than 3$R_{e}$. They find $(M/L)_{R_c}$=10.0, as well as concluding that mass follows light in this object. Our inferred IMFs (from an integrated spectrum out to 0.65 $R_e$), combined with published age and metallicity measurements, lead to a final value of $(M/L)_{R_c}=5.04\pm2.26$\footnote{Found by combining the values $(M/L)_{R_c}=4.2^{+0.6}_{- 0.5}$ from spectral fitting, $8.09^{+4.1}_{- 2.3}$ using the E-MILES models and $6.0^{+2.8}_{- 1.6}$ using the CvD12 models}.  This is lower than the value from \citeauthor{Thomas}, despite the fact that the dynamics in this object were fit without a dark matter halo term (i.e with mass following light). This may be evidence, therefore, for a dark matter profile which closely follows the visible matter in this object.

\section{Discussion}
\label{sec:Discussion}

The main result of this work is the strong gradient in NaI0.82 absorption combined with flat profiles for FeH0.99 in both objects. The equivalent widths of FeH in both galaxies also scatter around a similar value: 0.4 \AA~at a velocity dispersion of 200\kms. However, whilst the FeH index values are similar between the galaxies, the measured global IMF from full spectral fitting are significantly different: $x=2.69^{+0.11}_{-0.11}$ for NGC~1277 and $x=2.27^{+0.16}_{-0.18}$ for IC~843.

This may imply that relying on the Wing-Ford band alone to determine the single-power law IMF slope in an object could lead to different results than when utilising information from a number of gravity sensitive indices at once. However, regardless of the method used to infer the low-mass IMF slope in these objects, both these galaxies show evidence for an IMF significantly different from the IMF in the Milky Way, in agreement with previous work finding evidence for a non-universal IMF in massive early-type galaxies.

Recent work has called into question the efficacy of the Wing-Ford band as a reliable IMF indicator.  Using a two-part power law to characterise the low-mass IMF, \cite{2017ApJ...841...68V} show that the FeH equivalent width does not correlate with the IMF "mismatch" parameter, $\alpha$, in their study, with some of the weakest FeH index measurements in galaxies with very bottom heavy IMFs. Furthermore, \cite{LaBarbera2016} show that unimodal IMF determinations from the Wing-Ford band are in tension with unimodal IMF measurements from optical IMF sensitive indices (and use this fact to constrain the shape of the IMF in this galaxy). 

\subsection{Resolved IMF inferences}
With the wavelength range used in this work, we are unable to reliably constrain some of the important stellar population parameters necessary to infer quantitative radial IMF measurements and disentangle the effects of IMF variation from abundance gradients. We do, however, present qualitative discussion of the radial trends implied by our measurements. We find that IMF gradients by themselves, with no variations in the abundances of [Na/Fe] or [Fe/H] are only marginally consistent with our radial FeH and NaI index measurements. Plausible gradients in these elemental abundances, combined with a radially constant IMF, are more consistent with the data from NGC~1277 and IC~843.

\begin{figure*}
\centering
\includegraphics[width=\linewidth]{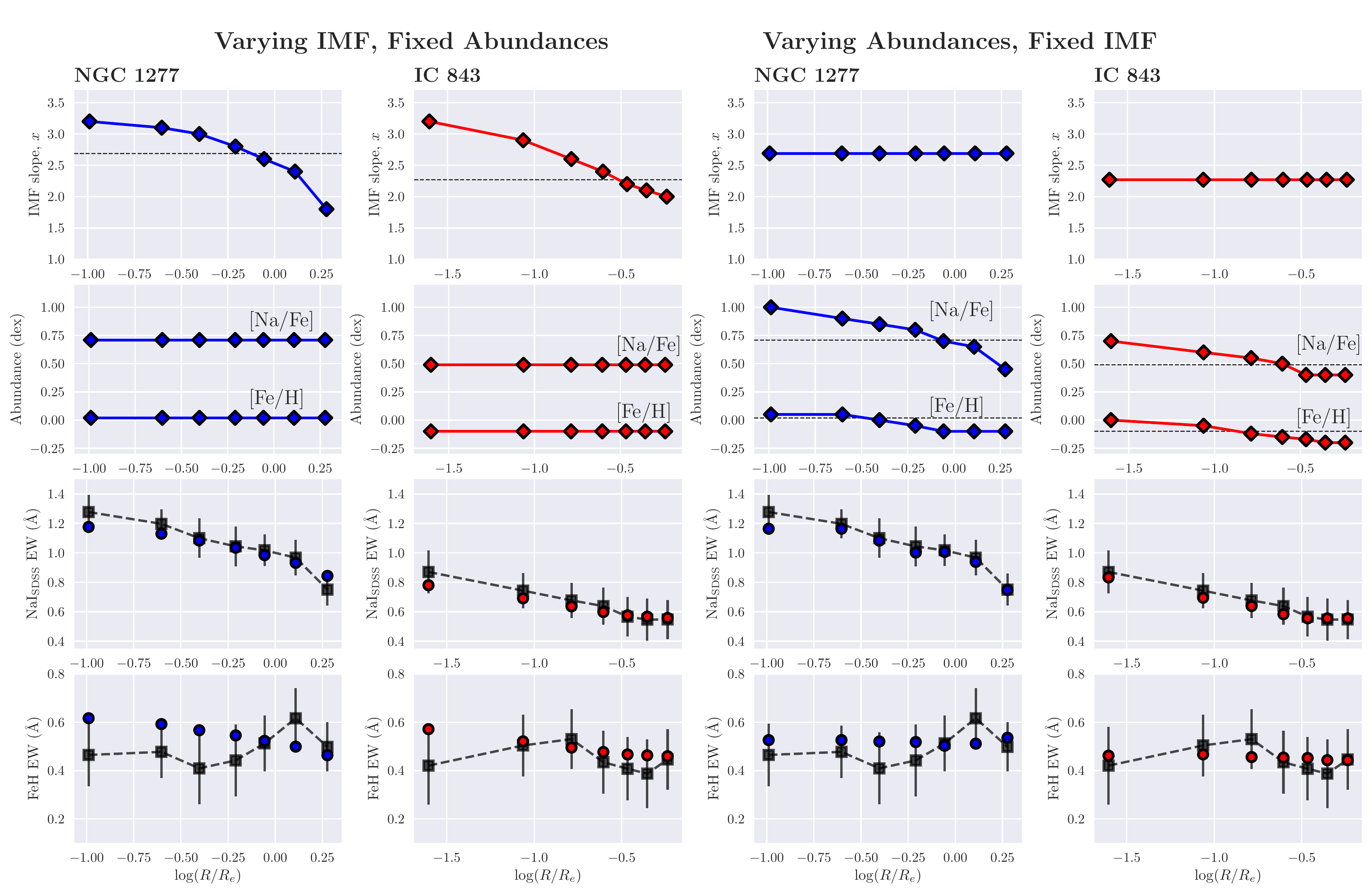}
\caption{A comparison of our index measurements as a function of radius to the CvD12 models. We create model spectra to correspond to the seven radial index measurements we make in each galaxy. Each model spectrum is convolved to 200\kms, in order to match the index measurements. We assume a constant age and [$\alpha$/Fe] for each spectrum, then impose either a radially varying IMF and fixed abundance patterns in [Na/Fe] and [Fe/H] (left two columns) or a radially constant IMF combined with abundance gradients (right two columns).  Dotted lines show the abundance values and IMF slope found from full spectral fitting in each galaxy. The third and fourth rows show both our measurements (black squares) and these model index measurements (coloured circles) for the NaI and FeH indices. Note that this is not a fit to these data, and we do not aim to draw quantitative conclusions about these abundance or IMF gradients. We find that a radially constant IMF is marginally consistent with our measurements, whilst radial gradients in [Na/Fe] and [Fe/H] are more consistent with the data. It is likely that a combination of IMF and abundance gradients exist in NGC~1277 and IC~843, although we are unable to break this degeneracy and quantify their magnitudes.}
\label{fig:exmple_flat+IMF_model}
\end{figure*}

Figure \ref{fig:exmple_flat+IMF_model} motivates this conclusion. We have produced mock spectra from the CvD12 models with varying values of [Na/Fe], [Fe/H] and low-mass IMF slope, $x$, all convolved to 200\kms. We measure the FeH and NaI indices from these spectra and compare to our index measurements. The top row of Figure 8 shows the IMF slope for each mock spectrum. The second row shows the assumed [Na/Fe] and [Fe/H] abundances. Rows 3 and 4 shows the NaI and FeH indices from the mock spectra, as well as our measurements from NGC~1277 and IC~843. Each panel is plotted against radial position. 

In the left two columns, we vary the IMF slope as a function of radius and fix the values of [Na/Fe] and [Fe/H] to those found from spectral fitting in each galaxy. The right two columns show a fixed IMF slope (again, fixed to the values measured using spectral fitting) and vary the abundances of [Na/Fe] and [Fe/H]. In all cases, we assume a constant age of 13 Gyr for NGC~1277 and 10 Gyr for IC~843, as well as an [$\alpha$/Fe] abundance of +0.25 dex for both galaxies. 

We note that this is not a fit to the data; we do not attempt to minimise a $\chi^{2}$-like function, or infer quantitative values of abundance gradients from this process. We simply vary the assumed IMF and abundance values to best recover the observed measurements. We also note that this exercise is vastly simplified, since we are fixing the values of stellar age, metallicity and [$\alpha$/Fe] to be held constant, although adding in further complexity would only further increase the degeneracies noted here.

When fixing the chemical abundances and varying the IMF as a function of radius, our mock spectra tend to under predict the NaI index whilst over-predicting FeH in the centre of both galaxies, although both are still consistent at the edge of the error bars. On the other hand, a flat IMF with plausible abundance gradients seems to better match the data, with the very central value of NaI in NGC~1277 the only place where the model and measurements are marginally in tension. Whilst the absolute values of the [Na/Fe] abundance needed to match the NaI measurements in NGC~1277 are very large, we note that the overall gradient of $\Delta$[Na/Fe]$=\sim-0.3$ dex per decade in $\log(R/R_e)$is plausible \citep[e.g.][]{2017MNRAS.468.1594A}. We also note there is a different response to [Na/Fe] overabundances in the CvD12 and E-MILES models, implying that the absolute values of [Na/Fe] abundance needed to match these measurements is likely to be model dependent.

Gradients in [Na/Fe] within individual galaxies have been recently been measured in the context of IMF variations. \cite{2017ApJ...841...68V} used long-slit spectroscopy and full-spectral fitting to measure abundance gradients in six nearby ETGs as well as gradients in the IMF. Furthermore,  \cite{2017MNRAS.468.1594A} also measure a gradient in [Na/Fe] in a stack of 8 nearby ETGs, finding $\Delta$[Na/Fe]=-0.35 dex per decade in $\log(R/R_e)$. Similar abundance gradients were measured for those individual galaxies in the stack with high enough quality data. Interestingly, unlike \cite{2017ApJ...841...68V}, they find no evidence for IMF gradients in their data, with the IMF in their stacked spectrum being uniformly Salpeter throughout.  

Super solar [Na/Fe] abundance ratios are also not uncommon in massive ETGs. \cite{2013ApJS..208....7J} find excess NaD line strengths in $\sim$8\% of low redshift ($z<0.08$) SDSS DR7 galaxies, including in ETGs without visible dust lanes, and conclude that [Na/Fe] enhancement, rather than ISM or IMF effects, are the cause. Furthermore, both \cite{2014ApJ...783...20W} and \cite{2014ApJ...780...33C} find a trend of increasing [Na/Fe] abundance in galaxies with larger velocity dispersions, of up to $\sim 0.4$ dex in galaxies with $\sigma$=300\kms, using independent SPS models. 

Abundance gradients and IMF gradients are not mutually exclusive, of course, and it is very plausible that a mixture of both exist in NGC~1277 and IC~843. These quantities are notoriously difficult to disentangle, and we would require coverage of a greater number of gravity sensitive absorption features to break the degeneracy and make quantitative statements in these objects.

A key assumption in this work is that the low-mass IMF slope in these objects is a single power law. A further explanation of our measurements could be that the IMF varies radially but does not have such a shape. In the "bimodal" parameterisation of \cite{1996ApJS..106..307V}, the IMF is flat at masses below 0.2 M$_{\odot}$ whilst the high mass slope (above 0.6 M$_{\odot}$) varies. The region in between is connected by spline interpolation. Such an IMF shape introduces a degeneracy between the FeH and NaI indices, by decoupling the very low-mass end of the IMF (where FeH is most sensitive) from the region between $0.2 M_{\odot} < M < 0.6 M_{\odot}$ (where NaI is most sensitive). Qualitatively, this allows a change in NaI strength without the corresponding change in FeH.  \cite{LaBarbera2016} use this form of the IMF in their study of a nearby massive ETG, finding radially constant measurements of FeH as well as a bimodal IMF gradient. MN15 also use this bimodal IMF parameterisation in their study of NGC~1277. They found evidence for a bottom heavy bimodal IMF of $\Gamma_b=3$ out to 0.3 $R_{e}$, which decreases and flattens off to  $\Gamma_b=2.5$ between 0.8 R$_e$ and 1.4 R$_{e}$.  

Furthermore, \cite{CvD12} and \cite{2017ApJ...841...68V} also use a parameterisation of the IMF which is not a single power law. Their IMF is fixed at high masses (above 1 M$_{\odot}$) and is a two part power law below, with a break at 0.5 M$_{\odot}$. This form of the low-mass IMF would again allow the NaI and FeH indices to vary independently of each other too, and could explain their behaviour in NGC~1277 and IC~843. Under this form of the IMF, \cite{2017ApJ...841...68V} explicitly show that a lack of gradient in FeH does not imply a radially constant IMF. 

With the data available to us, we are unable to rule out the possibility that our NaI and FeH measurements are caused by chemical abundance gradients, radial IMF gradients, or a combination of the two. A wavelength range covering further spectral indices, such as the Na D index and various optical Fe lines,  in conjunction with the publicly available state of the art stellar population models, would allow us to make a clearer separation of the effects of the low-mass IMF from abundance gradients.

\subsection{Other radial studies of FeH and NaI indices}
\label{sec:other_studies}

Similar measurements to ours have been found by other authors who have investigated both NaI and FeH indices as a function of radius in a variety of objects. As mentioned previously, \cite{2017MNRAS.468.1594A} find a strong gradient in NaI and radially flat FeH in a stack of 8 massive ETGs. They find uniform, typically bottom heavy, IMFs in their stack and most of their individual galaxies, with radial changes in index strengths primarily accounted for by abundance gradients. \cite{Z15} studied the central bulge of M31, observing a large decrease in NaI combined with no radial change in FeH, and conclude in favour of a gradient in [Na/Fe] rather than the IMF. Furthermore, \cite{2017MNRAS.465..192Z} studied the brightest cluster galaxies (BCGs) in the Coma cluster, measuring a strong gradient in NaI combined with flat FeH profile in the massive ETG NGC~4889, which has a central velocity dispersion of nearly 400\kms. Other objects in the sample also show weak FeH absorption. Only NGC~4839 displays evidence for a deep Wing-Ford band, although large systematic uncertainty due to residual sky emission prevents the authors from drawing strong conclusions about its stellar population.

\citet{McConnell2016} obtained deep long-slit data on two nearby ETGs, both of which had been part of the \cite{CvD12} sample. They found strong gradients in NaI but a much weaker decline in FeH, as well as opposite behaviour in NaI/$\langle$Fe$\rangle$ and FeH/$\langle$Fe$\rangle$. Again, the authors conclude in favour of a variation in [Na/Fe] over the central $\sim$300 pc of each galaxy instead of variation in a single power law IMF driving the strong decline in NaI. The authors also argue that the flat FeH profile implies a fixed low-mass slope of the IMF below M $\lesssim0.4M_{\odot}$.

Finally, \cite{LaBarbera2016, 2017MNRAS.464.3597L} make resolved measurements of the Wing-Ford band and a number of Na indices to constrain the shape of the low-mass IMF in a nearby ETG. They too find a lack of radial variation in FeH combined with negative gradients in NaI, NaD and two further Na lines at 1.14 and 2.21 $\mu$m, from which they find a gradient in a bimodal IMF combined with a radial change in [Na/Fe].

\section{Conclusions}
\label{sec:Conclusion}

We have used the Oxford SWIFT instrument to undertake a study of two low redshift early-type galaxies in order to make resolved measurements IMF sensitive indices in their spectra. We obtained high S/N integral field data of NGC~1277, a fast rotator in the Perseus cluster with a very high central velocity dispersion, and IC~843, also a fast rotator, located in the Coma cluster. Our measurements extend radially to 7.7\arcsec~and 6.2\arcsec~respectively, corresponding to 2.2 and 0.65 $R_{e}$. The SWIFT wavelength coverage, from 6300 \AA~to 10412 \AA, allows resolved measurement of the NaI doublet, CaII triplet, TiO, MgI and FeH spectral features. We conclude:

\begin{enumerate}

\item NGC~1277 shows a strong negative gradient in NaI, more marginal negative TiO and CaT gradients and a flat FeH profile. The FeH equivalent widths scatter around 0.42 \AA~at all radii (corrected to 200\kms). 

\item IC~843 is similar, if less extreme, than NGC~1277. It displays a weaker NaI and TiO gradients, and flat profiles in FeH, CaT and MgI. FeH equivalent widths are a similar strength to NGC~1277, also around 0.4 \AA. 

\item Similarly to \citet{McConnell2016}, \cite{2017MNRAS.465..192Z}, \cite{2017MNRAS.468.1594A} and others, we find very different radial trends between the IMF sensitive indices NaI and FeH. 

\item In both NGC~1277 and IC~843, our measurements can be explained by a radially constant single power law IMF combined with appropriate abundance gradients. However, a radial gradient in the IMF may also reproduce our results, and our data do not allow us to break this degeneracy, as shown in Figure 8. Furthermore, with the spectral range available from SWIFT, gradients in more complicated IMF parameterisations (such as a bimodal or multi-segment IMF) also cannot be excluded. A wavelength range covering absorption indices sensitive to stellar population parameters such as age, [Z/H] and [$\alpha/$Fe], as well as indices sensitive to elemental abundances (such as NaD and the combination of Fe5250 and Fe5335) are vital to isolate the effects of the IMF. 

\item We use our global spectra and state-of-the-art stellar population models to infer global single power-law IMFs in each object. We determine the IMF in each object using three techniques: full spectral fitting using the CvD12 models, as well as fitting the FeH index with corrections for non-solar abundance patterns using the CvD12 and E-MILES models \citep[following][]{2017MNRAS.465..192Z}. We find that a super Salpeter slope fits best in NGC~1277, with each technique in agreement. IC~843 is more uncertain, with the spectral fitting consistent with a Salpeter IMF and the index fitting scattering higher.  

\item Despite NGC~1277 and IC~843 having only a $\lesssim10\%$ difference in global FeH measurement, and similar population parameters, we find significantly different global IMF slopes when we include further spectral information from between 6600\AA~and 10,000\AA. This may bring into doubt the use of Wing-Ford band to infer an IMF slope when not combined with information from other areas of the spectrum.

\item Our inferred V band stellar mass-to-light ratios are in agreement with published dynamical and spectroscopic determinations. In IC~843, we find a mass-to-light ratio ($R_c$ band) smaller than the dynamical $(M/L)$ from \cite{Thomas}, despite their conclusion that the dynamics can be fit without a dark matter halo term (i.e that mass follows light). This may imply a non-standard dark matter profile in this object.

\end{enumerate}

\section*{Acknowledgements}

We would like to thank F. La Barbera for a referee report which greatly improved this work, as well as for making available the Na-MILES models used in this paper. SPV would like to thank P. Alton for fruitful discussions on the effect of Na on stellar atmospheres, and A. Ferr\'{e}-Mateu for making her optical measurements of NGC~1277 available.

This paper made use of the Astropy python package \citep{astropy}, as well as the \texttt{matplotlib} \citep{matplotlib} and \texttt{seaborn} plotting software \citep{seaborn} and the scientific libraries \texttt{numpy} \citep{numpy} and \texttt{scipy} \citep{scipy}. 

The Oxford SWIFT integral field spectrograph was supported by a Marie Curie Excellence Grant from the European Commission (MEXT-CT-2003-002792, Team Leader: N. Thatte). It was also supported by additional funds from the University of Oxford Physics Department and the John Fell OUP Research Fund. Additional funds to host and support SWIFT at the 200-inch Hale Telescope on Palomar were provided by Caltech Optical Observatories. 

This work was supported by the Astrophysics at Oxford grants (ST/H002456/1 and ST/K00106X/1) as well as visitors grant (ST/H504862/1) from the UK Science and Technology Facilities Council. SPV is supported by a doctoral studentship supported by STFC grant ST/N504233/1. RCWH was supported by the Science and Technology Facilities Council (STFC grant numbers ST/H002456/1, ST/K00106X/1 \& ST/J002216/1). RLD acknowledges travel and computer grants from Christ
Church, Oxford, and support from the Oxford Hintze Centre for Astrophysical Surveys, which is funded through generous support from the Hintze Family Charitable Foundation.

\bibliographystyle{mnras}
\bibliography{ms}

\begin{thebibliography}{}
\makeatletter
\relax
\def\mn@urlcharsother{\let\do\@makeother \do\$\do\&\do\#\do\^\do\_\do\%\do\~}
\def\mn@doi{\begingroup\mn@urlcharsother \@ifnextchar [ {\mn@doi@}
  {\mn@doi@[]}}
\def\mn@doi@[#1]#2{\def\@tempa{#1}\ifx\@tempa\@empty \href
  {http://dx.doi.org/#2} {doi:#2}\else \href {http://dx.doi.org/#2} {#1}\fi
  \endgroup}
\def\mn@eprint#1#2{\mn@eprint@#1:#2::\@nil}
\def\mn@eprint@arXiv#1{\href {http://arxiv.org/abs/#1} {{\tt arXiv:#1}}}
\def\mn@eprint@dblp#1{\href {http://dblp.uni-trier.de/rec/bibtex/#1.xml}
  {dblp:#1}}
\def\mn@eprint@#1:#2:#3:#4\@nil{\def\@tempa {#1}\def\@tempb {#2}\def\@tempc
  {#3}\ifx \@tempc \@empty \let \@tempc \@tempb \let \@tempb \@tempa \fi \ifx
  \@tempb \@empty \def\@tempb {arXiv}\fi \@ifundefined
  {mn@eprint@\@tempb}{\@tempb:\@tempc}{\expandafter \expandafter \csname
  mn@eprint@\@tempb\endcsname \expandafter{\@tempc}}}

\bibitem[\protect\citeauthoryear{{Alton}, {Smith}  \& {Lucey}}{{Alton}
  et~al.}{2017}]{2017MNRAS.468.1594A}
{Alton} P.~D.,  {Smith} R.~J.,   {Lucey} J.~R.,  2017, \mn@doi [\mnras]
  {10.1093/mnras/stx464}, \href
  {http://adsabs.harvard.edu/abs/2017MNRAS.468.1594A} {468, 1594}

\bibitem[\protect\citeauthoryear{{Anderson} \& {Darling}}{{Anderson} \&
  {Darling}}{1954}]{AndersonDarling}
{Anderson} T.~W.,  {Darling} D.~A.,  1954, \mn@doi [Journal of the American
  Statistical Association] {10.1080/01621459.1954.10501232}, 49, 765

\bibitem[\protect\citeauthoryear{{Astropy Collaboration} et~al.,}{{Astropy
  Collaboration} et~al.}{2013}]{astropy}
{Astropy Collaboration} et~al., 2013, \mn@doi [\aap]
  {10.1051/0004-6361/201322068}, \href
  {http://adsabs.harvard.edu/abs/2013A%26A...558A..33A} {558, A33}

\bibitem[\protect\citeauthoryear{{Barnab{\`e}}, {Spiniello}, {Koopmans},
  {Trager}, {Czoske}  \& {Treu}}{{Barnab{\`e}}
  et~al.}{2013}]{2013MNRAS.436..253B}
{Barnab{\`e}} M.,  {Spiniello} C.,  {Koopmans} L.~V.~E.,  {Trager} S.~C.,
  {Czoske} O.,   {Treu} T.,  2013, \mn@doi [\mnras] {10.1093/mnras/stt1727},
  \href {http://adsabs.harvard.edu/abs/2013MNRAS.436..253B} {436, 253}

\bibitem[\protect\citeauthoryear{{Bastian}, {Covey}  \& {Meyer}}{{Bastian}
  et~al.}{2010}]{Bastian}
{Bastian} N.,  {Covey} K.~R.,   {Meyer} M.~R.,  2010, \mn@doi [\araa]
  {10.1146/annurev-astro-082708-101642}, \href
  {http://adsabs.harvard.edu/abs/2010ARA%26A..48..339B} {48, 339}

\bibitem[\protect\citeauthoryear{Cappellari \& Emsellem}{Cappellari \&
  Emsellem}{2004}]{ppxf}
Cappellari M.,  Emsellem E.,  2004, Publications of the Astronomical Society of
  the Pacific, 116, 138

\bibitem[\protect\citeauthoryear{{Cappellari} et~al.,}{{Cappellari}
  et~al.}{2011}]{2011MNRAS.413..813C}
{Cappellari} M.,  et~al., 2011, \mn@doi [\mnras]
  {10.1111/j.1365-2966.2010.18174.x}, \href
  {http://adsabs.harvard.edu/abs/2011MNRAS.413..813C} {413, 813}

\bibitem[\protect\citeauthoryear{{Cappellari} et~al.,}{{Cappellari}
  et~al.}{2013}]{2013MNRAS.432.1862C}
{Cappellari} M.,  et~al., 2013, \mn@doi [\mnras] {10.1093/mnras/stt644}, \href
  {http://adsabs.harvard.edu/abs/2013MNRAS.432.1862C} {432, 1862}

\bibitem[\protect\citeauthoryear{{Cenarro}, {Cardiel}, {Gorgas}, {Peletier},
  {Vazdekis}  \& {Prada}}{{Cenarro} et~al.}{2001}]{Cenarro2001}
{Cenarro} A.~J.,  {Cardiel} N.,  {Gorgas} J.,  {Peletier} R.~F.,  {Vazdekis}
  A.,   {Prada} F.,  2001, \mn@doi [\mnras] {10.1046/j.1365-8711.2001.04688.x},
  \href {http://adsabs.harvard.edu/abs/2001MNRAS.326..959C} {326, 959}

\bibitem[\protect\citeauthoryear{{Cenarro}, {Gorgas}, {Vazdekis}, {Cardiel}  \&
  {Peletier}}{{Cenarro} et~al.}{2003}]{2003MNRAS.339L..12C}
{Cenarro} A.~J.,  {Gorgas} J.,  {Vazdekis} A.,  {Cardiel} N.,   {Peletier}
  R.~F.,  2003, \mn@doi [\mnras] {10.1046/j.1365-8711.2003.06360.x}, \href
  {http://adsabs.harvard.edu/abs/2003MNRAS.339L..12C} {339, L12}

\bibitem[\protect\citeauthoryear{{Chabrier}}{{Chabrier}}{2003}]{Chabrier}
{Chabrier} G.,  2003, \mn@doi [\pasp] {10.1086/376392}, \href
  {http://adsabs.harvard.edu/abs/2003PASP..115..763C} {115, 763}

\bibitem[\protect\citeauthoryear{{Clauwens}, {Schaye}  \& {Franx}}{{Clauwens}
  et~al.}{2016}]{2016MNRAS.462.2832C}
{Clauwens} B.,  {Schaye} J.,   {Franx} M.,  2016, \mn@doi [\mnras]
  {10.1093/mnras/stw1808}, \href
  {http://adsabs.harvard.edu/abs/2016MNRAS.462.2832C} {462, 2832}

\bibitem[\protect\citeauthoryear{{Clough}, {Shephard}, {Mlawer}, {Delamere},
  {Iacono}, {Cady-Pereira}, {Boukabara}  \& {Brown}}{{Clough}
  et~al.}{2005}]{Clough2005}
{Clough} S.~A.,  {Shephard} M.~W.,  {Mlawer} E.~J.,  {Delamere} J.~S.,
  {Iacono} M.~J.,  {Cady-Pereira} K.,  {Boukabara} S.,   {Brown} P.~D.,  2005,
  \mn@doi [\jqsrt] {10.1016/j.jqsrt.2004.05.058}, \href
  {http://adsabs.harvard.edu/abs/2005JQSRT..91..233C} {91, 233}

\bibitem[\protect\citeauthoryear{{Cohen}}{{Cohen}}{1978}]{1978ApJ...221..788C}
{Cohen} J.~G.,  1978, \mn@doi [\apj] {10.1086/156081}, \href
  {http://adsabs.harvard.edu/abs/1978ApJ...221..788C} {221, 788}

\bibitem[\protect\citeauthoryear{{Conroy} \& {van Dokkum}}{{Conroy} \& {van
  Dokkum}}{2012}]{CvD12a}
{Conroy} C.,  {van Dokkum} P.,  2012, \mn@doi [\apj]
  {10.1088/0004-637X/747/1/69}, \href
  {http://adsabs.harvard.edu/abs/2012ApJ...747...69C} {747, 69}

\bibitem[\protect\citeauthoryear{{Conroy}, {Graves}  \& {van Dokkum}}{{Conroy}
  et~al.}{2014}]{2014ApJ...780...33C}
{Conroy} C.,  {Graves} G.~J.,   {van Dokkum} P.~G.,  2014, \mn@doi [\apj]
  {10.1088/0004-637X/780/1/33}, \href
  {http://adsabs.harvard.edu/abs/2014ApJ...780...33C} {780, 33}

\bibitem[\protect\citeauthoryear{{Couture} \& {Hardy}}{{Couture} \&
  {Hardy}}{1993}]{Couture_Hardy}
{Couture} J.,  {Hardy} E.,  1993, \mn@doi [\apj] {10.1086/172426}, \href
  {http://adsabs.harvard.edu/abs/1993ApJ...406..142C} {406, 142}

\bibitem[\protect\citeauthoryear{{Davies}}{{Davies}}{2007}]{2007MNRAS.375.1099D}
{Davies} R.~I.,  2007, \mn@doi [\mnras] {10.1111/j.1365-2966.2006.11383.x},
  \href {http://adsabs.harvard.edu/abs/2007MNRAS.375.1099D} {375, 1099}

\bibitem[\protect\citeauthoryear{{Davies}, {Sadler}  \& {Peletier}}{{Davies}
  et~al.}{1993}]{1993MNRAS.262..650D}
{Davies} R.~L.,  {Sadler} E.~M.,   {Peletier} R.~F.,  1993, \mn@doi [\mnras]
  {10.1093/mnras/262.3.650}, \href
  {http://adsabs.harvard.edu/abs/1993MNRAS.262..650D} {262, 650}

\bibitem[\protect\citeauthoryear{{Emsellem}}{{Emsellem}}{2013}]{2013MNRAS.433.1862E}
{Emsellem} E.,  2013, \mn@doi [\mnras] {10.1093/mnras/stt840}, \href
  {http://adsabs.harvard.edu/abs/2013MNRAS.433.1862E} {433, 1862}

\bibitem[\protect\citeauthoryear{{Faber}}{{Faber}}{1983}]{1983HiA.....6..165F}
{Faber} S.~M.,  1983, Highlights of Astronomy, \href
  {http://adsabs.harvard.edu/abs/1983HiA.....6..165F} {6, 165}

\bibitem[\protect\citeauthoryear{{Faber} \& {French}}{{Faber} \&
  {French}}{1980}]{1980LicOB.823....1F}
{Faber} S.~M.,  {French} H.~B.,  1980, Lick Observatory Bulletin, \href
  {http://adsabs.harvard.edu/abs/1980LicOB.823....1F} {823, 1}

\bibitem[\protect\citeauthoryear{{Ferr{\'e}-Mateu}, {Trujillo},
  {Mart{\'{\i}}n-Navarro}, {Vazdekis}, {Mezcua}, {Balcells}  \&
  {Dom{\'{\i}}nguez}}{{Ferr{\'e}-Mateu} et~al.}{2017}]{2017MNRAS.tmp..177F}
{Ferr{\'e}-Mateu} A.,  {Trujillo} I.,  {Mart{\'{\i}}n-Navarro} I.,  {Vazdekis}
  A.,  {Mezcua} M.,  {Balcells} M.,   {Dom{\'{\i}}nguez} L.,  2017, \mn@doi
  [\mnras] {10.1093/mnras/stx171}, \href
  {http://adsabs.harvard.edu/abs/2017MNRAS.tmp..177F} {}

\bibitem[\protect\citeauthoryear{{Ferreras}, {La Barbera}, {de la Rosa},
  {Vazdekis}, {de Carvalho}, {Falc{\'o}n-Barroso}  \&
  {Ricciardelli}}{{Ferreras} et~al.}{2013}]{2013MNRAS.429L..15F}
{Ferreras} I.,  {La Barbera} F.,  {de la Rosa} I.~G.,  {Vazdekis} A.,  {de
  Carvalho} R.~R.,  {Falc{\'o}n-Barroso} J.,   {Ricciardelli} E.,  2013,
  \mn@doi [\mnras] {10.1093/mnrasl/sls014}, \href
  {http://adsabs.harvard.edu/abs/2013MNRAS.429L..15F} {429, L15}

\bibitem[\protect\citeauthoryear{{Foreman-Mackey}, {Hogg}, {Lang}  \&
  {Goodman}}{{Foreman-Mackey} et~al.}{2013}]{2013PASP..125..306F}
{Foreman-Mackey} D.,  {Hogg} D.~W.,  {Lang} D.,   {Goodman} J.,  2013, \mn@doi
  [\pasp] {10.1086/670067}, \href
  {http://adsabs.harvard.edu/abs/2013PASP..125..306F} {125, 306}

\bibitem[\protect\citeauthoryear{{Hopkins}, {Bundy}, {Murray}, {Quataert},
  {Lauer}  \& {Ma}}{{Hopkins} et~al.}{2009}]{2009MNRAS.398..898H}
{Hopkins} P.~F.,  {Bundy} K.,  {Murray} N.,  {Quataert} E.,  {Lauer} T.~R.,
  {Ma} C.-P.,  2009, \mn@doi [\mnras] {10.1111/j.1365-2966.2009.15062.x}, \href
  {http://adsabs.harvard.edu/abs/2009MNRAS.398..898H} {398, 898}

\bibitem[\protect\citeauthoryear{Hunter}{Hunter}{2007}]{matplotlib}
Hunter J.~D.,  2007, \mn@doi [Computing In Science \& Engineering]
  {10.1109/MCSE.2007.55}, 9, 90

\bibitem[\protect\citeauthoryear{{Jeong}, {Yi}, {Kyeong}, {Sarzi}, {Sung}  \&
  {Oh}}{{Jeong} et~al.}{2013}]{2013ApJS..208....7J}
{Jeong} H.,  {Yi} S.~K.,  {Kyeong} J.,  {Sarzi} M.,  {Sung} E.-C.,   {Oh} K.,
  2013, \mn@doi [\apjs] {10.1088/0067-0049/208/1/7}, \href
  {http://adsabs.harvard.edu/abs/2013ApJS..208....7J} {208, 7}

\bibitem[\protect\citeauthoryear{Jones, Oliphant, Peterson  et~al.}{Jones
  et~al.}{2001}]{scipy}
Jones E.,  Oliphant T.,  Peterson P.,   et~al., 2001, {SciPy}: Open source
  scientific tools for {Python}, \url {http://www.scipy.org/}

\bibitem[\protect\citeauthoryear{{Kausch} et~al.,}{{Kausch}
  et~al.}{2014}]{Molecfit}
{Kausch} W.,  et~al., 2014, in {Manset} N.,  {Forshay} P.,  eds,  Astronomical
  Society of the Pacific Conference Series Vol. 485, Astronomical Data Analysis
  Software and Systems XXIII. p.~403 (\mn@eprint {arXiv} {1401.7768})

\bibitem[\protect\citeauthoryear{{Kroupa}}{{Kroupa}}{2001}]{Kroupa}
{Kroupa} P.,  2001, \mn@doi [\mnras] {10.1046/j.1365-8711.2001.04022.x}, \href
  {http://adsabs.harvard.edu/abs/2001MNRAS.322..231K} {322, 231}

\bibitem[\protect\citeauthoryear{{La Barbera}, {Ferreras}, {Vazdekis}, {de la
  Rosa}, {de Carvalho}, {Trevisan}, {Falc{\'o}n-Barroso}  \&
  {Ricciardelli}}{{La Barbera} et~al.}{2013a}]{LaBarbera2013}
{La Barbera} F.,  {Ferreras} I.,  {Vazdekis} A.,  {de la Rosa} I.~G.,  {de
  Carvalho} R.~R.,  {Trevisan} M.,  {Falc{\'o}n-Barroso} J.,   {Ricciardelli}
  E.,  2013a, \mn@doi [\mnras] {10.1093/mnras/stt943}, \href
  {http://adsabs.harvard.edu/abs/2013MNRAS.433.3017L} {433, 3017}

\bibitem[\protect\citeauthoryear{{La Barbera}, {Ferreras}, {Vazdekis}, {de la
  Rosa}, {de Carvalho}, {Trevisan}, {Falc{\'o}n-Barroso}  \&
  {Ricciardelli}}{{La Barbera} et~al.}{2013b}]{2013MNRAS.433.3017L}
{La Barbera} F.,  {Ferreras} I.,  {Vazdekis} A.,  {de la Rosa} I.~G.,  {de
  Carvalho} R.~R.,  {Trevisan} M.,  {Falc{\'o}n-Barroso} J.,   {Ricciardelli}
  E.,  2013b, \mn@doi [\mnras] {10.1093/mnras/stt943}, \href
  {http://adsabs.harvard.edu/abs/2013MNRAS.433.3017L} {433, 3017}

\bibitem[\protect\citeauthoryear{{La Barbera}, {Vazdekis}, {Ferreras},
  {Pasquali}, {Cappellari}, {Mart{\'{\i}}n-Navarro}, {Sch{\"o}nebeck}  \&
  {Falc{\'o}n-Barroso}}{{La Barbera} et~al.}{2016}]{LaBarbera2016}
{La Barbera} F.,  {Vazdekis} A.,  {Ferreras} I.,  {Pasquali} A.,  {Cappellari}
  M.,  {Mart{\'{\i}}n-Navarro} I.,  {Sch{\"o}nebeck} F.,   {Falc{\'o}n-Barroso}
  J.,  2016, \mn@doi [\mnras] {10.1093/mnras/stv2996}, \href
  {http://adsabs.harvard.edu/abs/2016MNRAS.457.1468L} {457, 1468}

\bibitem[\protect\citeauthoryear{{La Barbera}, {Vazdekis}, {Ferreras},
  {Pasquali}, {Allende Prieto}, {R{\"o}ck}, {Aguado}  \& {Peletier}}{{La
  Barbera} et~al.}{2017}]{2017MNRAS.464.3597L}
{La Barbera} F.,  {Vazdekis} A.,  {Ferreras} I.,  {Pasquali} A.,  {Allende
  Prieto} C.,  {R{\"o}ck} B.,  {Aguado} D.~S.,   {Peletier} R.~F.,  2017,
  \mn@doi [\mnras] {10.1093/mnras/stw2407}, \href
  {http://adsabs.harvard.edu/abs/2017MNRAS.464.3597L} {464, 3597}

\bibitem[\protect\citeauthoryear{{Lyubenova} et~al.,}{{Lyubenova}
  et~al.}{2016}]{2016MNRAS.463.3220L}
{Lyubenova} M.,  et~al., 2016, \mn@doi [\mnras] {10.1093/mnras/stw2434}, \href
  {http://adsabs.harvard.edu/abs/2016MNRAS.463.3220L} {463, 3220}

\bibitem[\protect\citeauthoryear{{Mart{\'{\i}}n-Navarro}, {Barbera},
  {Vazdekis}, {Falc{\'o}n-Barroso}  \& {Ferreras}}{{Mart{\'{\i}}n-Navarro}
  et~al.}{2015a}]{2015MNRAS.447.1033M}
{Mart{\'{\i}}n-Navarro} I.,  {Barbera} F.~L.,  {Vazdekis} A.,
  {Falc{\'o}n-Barroso} J.,   {Ferreras} I.,  2015a, \mn@doi [\mnras]
  {10.1093/mnras/stu2480}, \href
  {http://adsabs.harvard.edu/abs/2015MNRAS.447.1033M} {447, 1033}

\bibitem[\protect\citeauthoryear{{Mart{\'{\i}}n-Navarro}, {La Barbera},
  {Vazdekis}, {Ferr{\'e}-Mateu}, {Trujillo}  \&
  {Beasley}}{{Mart{\'{\i}}n-Navarro} et~al.}{2015b}]{MartinNavarro}
{Mart{\'{\i}}n-Navarro} I.,  {La Barbera} F.,  {Vazdekis} A.,
  {Ferr{\'e}-Mateu} A.,  {Trujillo} I.,   {Beasley} M.~A.,  2015b, \mn@doi
  [\mnras] {10.1093/mnras/stv1022}, \href
  {http://adsabs.harvard.edu/abs/2015MNRAS.451.1081M} {451, 1081}

\bibitem[\protect\citeauthoryear{{Mart{\'{\i}}n-Navarro}
  et~al.,}{{Mart{\'{\i}}n-Navarro} et~al.}{2015c}]{MartinNavarro2015c}
{Mart{\'{\i}}n-Navarro} I.,  et~al., 2015c, \mn@doi [\apjl]
  {10.1088/2041-8205/806/2/L31}, \href
  {http://adsabs.harvard.edu/abs/2015ApJ...806L..31M} {806, L31}

\bibitem[\protect\citeauthoryear{{McConnell}, {Lu}  \& {Mann}}{{McConnell}
  et~al.}{2016}]{McConnell2016}
{McConnell} N.~J.,  {Lu} J.~R.,   {Mann} A.~W.,  2016, \mn@doi [\apj]
  {10.3847/0004-637X/821/1/39}, \href
  {http://adsabs.harvard.edu/abs/2016ApJ...821...39M} {821, 39}

\bibitem[\protect\citeauthoryear{{Naab}, {Johansson}  \& {Ostriker}}{{Naab}
  et~al.}{2009}]{2009ApJ...699L.178N}
{Naab} T.,  {Johansson} P.~H.,   {Ostriker} J.~P.,  2009, \mn@doi [\apjl]
  {10.1088/0004-637X/699/2/L178}, \href
  {http://adsabs.harvard.edu/abs/2009ApJ...699L.178N} {699, L178}

\bibitem[\protect\citeauthoryear{{Price}, {Phillipps}, {Huxor}, {Smith}  \&
  {Lucey}}{{Price} et~al.}{2011}]{Price2011}
{Price} J.,  {Phillipps} S.,  {Huxor} A.,  {Smith} R.~J.,   {Lucey} J.~R.,
  2011, \mn@doi [\mnras] {10.1111/j.1365-2966.2010.17862.x}, \href
  {http://adsabs.harvard.edu/abs/2011MNRAS.411.2558P} {411, 2558}

\bibitem[\protect\citeauthoryear{{Salpeter}}{{Salpeter}}{1955}]{Salpeter}
{Salpeter} E.~E.,  1955, \mn@doi [\apj] {10.1086/145971}, \href
  {http://adsabs.harvard.edu/abs/1955ApJ...121..161S} {121, 161}

\bibitem[\protect\citeauthoryear{{Schiavon}, {Barbuy}, {Rossi}, {Milone}  \&
  {A.}}{{Schiavon} et~al.}{1997a}]{1997ApJ...479..902S}
{Schiavon} R.~P.,  {Barbuy} B.,  {Rossi} S.~C.~F.,  {Milone}  {A.} 1997a, \apj,
  \href {http://adsabs.harvard.edu/abs/1997ApJ...479..902S} {479, 902}

\bibitem[\protect\citeauthoryear{{Schiavon}, {Barbuy}  \& {Singh}}{{Schiavon}
  et~al.}{1997b}]{1997ApJ...484..499S}
{Schiavon} R.~P.,  {Barbuy} B.,   {Singh} P.~D.,  1997b, \apj, \href
  {http://adsabs.harvard.edu/abs/1997ApJ...484..499S} {484, 499}

\bibitem[\protect\citeauthoryear{{Smith}}{{Smith}}{2014}]{Smith2014}
{Smith} R.~J.,  2014, \mn@doi [\mnras] {10.1093/mnrasl/slu082}, \href
  {http://adsabs.harvard.edu/abs/2014MNRAS.443L..69S} {443, L69}

\bibitem[\protect\citeauthoryear{{Smith}, {Lucey}  \& {Conroy}}{{Smith}
  et~al.}{2015}]{2015MNRAS.449.3441S}
{Smith} R.~J.,  {Lucey} J.~R.,   {Conroy} C.,  2015, \mn@doi [\mnras]
  {10.1093/mnras/stv518}, \href
  {http://adsabs.harvard.edu/abs/2015MNRAS.449.3441S} {449, 3441}

\bibitem[\protect\citeauthoryear{{Spiniello}, {Trager}, {Koopmans}  \&
  {Conroy}}{{Spiniello} et~al.}{2014}]{2014MNRAS.438.1483S}
{Spiniello} C.,  {Trager} S.,  {Koopmans} L.~V.~E.,   {Conroy} C.,  2014,
  \mn@doi [\mnras] {10.1093/mnras/stt2282}, \href
  {http://adsabs.harvard.edu/abs/2014MNRAS.438.1483S} {438, 1483}

\bibitem[\protect\citeauthoryear{{Spiniello}, {Barnab{\`e}}, {Koopmans}  \&
  {Trager}}{{Spiniello} et~al.}{2015a}]{Spiniello2015a}
{Spiniello} C.,  {Barnab{\`e}} M.,  {Koopmans} L.~V.~E.,   {Trager} S.~C.,
  2015a, \mn@doi [\mnras] {10.1093/mnrasl/slv079}, \href
  {http://adsabs.harvard.edu/abs/2015MNRAS.452L..21S} {452, L21}

\bibitem[\protect\citeauthoryear{{Spiniello}, {Barnab{\`e}}, {Koopmans}  \&
  {Trager}}{{Spiniello} et~al.}{2015b}]{2015MNRAS.452L..21S}
{Spiniello} C.,  {Barnab{\`e}} M.,  {Koopmans} L.~V.~E.,   {Trager} S.~C.,
  2015b, \mn@doi [\mnras] {10.1093/mnrasl/slv079}, \href
  {http://adsabs.harvard.edu/abs/2015MNRAS.452L..21S} {452, L21}

\bibitem[\protect\citeauthoryear{{Spiniello}, {Trager}  \&
  {Koopmans}}{{Spiniello} et~al.}{2015c}]{2015ApJ...803...87S}
{Spiniello} C.,  {Trager} S.~C.,   {Koopmans} L.~V.~E.,  2015c, \mn@doi [\apj]
  {10.1088/0004-637X/803/2/87}, \href
  {http://adsabs.harvard.edu/abs/2015ApJ...803...87S} {803, 87}

\bibitem[\protect\citeauthoryear{{Spinrad} \& {Taylor}}{{Spinrad} \&
  {Taylor}}{1971}]{1971ApJS...22..445S}
{Spinrad} H.,  {Taylor} B.~J.,  1971, \mn@doi [\apjs] {10.1086/190232}, \href
  {http://adsabs.harvard.edu/abs/1971ApJS...22..445S} {22, 445}

\bibitem[\protect\citeauthoryear{{Thatte}, {Tecza}, {Clarke}, {Goodsall},
  {Lynn}, {Freeman}  \& {Davies}}{{Thatte} et~al.}{2006}]{2006SPIE.6269E..3LT}
{Thatte} N.,  {Tecza} M.,  {Clarke} F.,  {Goodsall} T.,  {Lynn} J.,  {Freeman}
  D.,   {Davies} R.~L.,  2006, in Society of Photo-Optical Instrumentation
  Engineers (SPIE) Conference Series. p. 62693L, \mn@doi{10.1117/12.670859}

\bibitem[\protect\citeauthoryear{{Thomas}, {Maraston}  \& {Bender}}{{Thomas}
  et~al.}{2003}]{2003MNRAS.343..279T}
{Thomas} D.,  {Maraston} C.,   {Bender} R.,  2003, \mn@doi [\mnras]
  {10.1046/j.1365-8711.2003.06659.x}, \href
  {http://adsabs.harvard.edu/abs/2003MNRAS.343..279T} {343, 279}

\bibitem[\protect\citeauthoryear{{Thomas}, {Saglia}, {Bender}, {Thomas},
  {Gebhardt}, {Magorrian}, {Corsini}  \& {Wegner}}{{Thomas}
  et~al.}{2007}]{Thomas}
{Thomas} J.,  {Saglia} R.~P.,  {Bender} R.,  {Thomas} D.,  {Gebhardt} K.,
  {Magorrian} J.,  {Corsini} E.~M.,   {Wegner} G.,  2007, \mn@doi [\mnras]
  {10.1111/j.1365-2966.2007.12434.x}, \href
  {http://adsabs.harvard.edu/abs/2007MNRAS.382..657T} {382, 657}

\bibitem[\protect\citeauthoryear{{Thomas}, {Maraston}  \& {Johansson}}{{Thomas}
  et~al.}{2011a}]{Thomas2011}
{Thomas} D.,  {Maraston} C.,   {Johansson} J.,  2011a, \mn@doi [\mnras]
  {10.1111/j.1365-2966.2010.18049.x}, \href
  {http://adsabs.harvard.edu/abs/2011MNRAS.412.2183T} {412, 2183}

\bibitem[\protect\citeauthoryear{{Thomas} et~al.,}{{Thomas}
  et~al.}{2011b}]{2011MNRAS.415..545T}
{Thomas} J.,  et~al., 2011b, \mn@doi [\mnras]
  {10.1111/j.1365-2966.2011.18725.x}, \href
  {http://adsabs.harvard.edu/abs/2011MNRAS.415..545T} {415, 545}

\bibitem[\protect\citeauthoryear{{Trager}, {Worthey}, {Faber}, {Burstein}  \&
  {Gonz{\'a}lez}}{{Trager} et~al.}{1998}]{1998ApJS..116....1T}
{Trager} S.~C.,  {Worthey} G.,  {Faber} S.~M.,  {Burstein} D.,   {Gonz{\'a}lez}
  J.~J.,  1998, \mn@doi [\apjs] {10.1086/313099}, \href
  {http://adsabs.harvard.edu/abs/1998ApJS..116....1T} {116, 1}

\bibitem[\protect\citeauthoryear{{Trager}, {Faber}, {Worthey}  \&
  {Gonz{\'a}lez}}{{Trager} et~al.}{2000}]{2000AJ....120..165T}
{Trager} S.~C.,  {Faber} S.~M.,  {Worthey} G.,   {Gonz{\'a}lez} J.~J.,  2000,
  \mn@doi [\aj] {10.1086/301442}, \href
  {http://adsabs.harvard.edu/abs/2000AJ....120..165T} {120, 165}

\bibitem[\protect\citeauthoryear{{Treu}, {Auger}, {Koopmans}, {Gavazzi},
  {Marshall}  \& {Bolton}}{{Treu} et~al.}{2010}]{2010ApJ...709.1195T}
{Treu} T.,  {Auger} M.~W.,  {Koopmans} L.~V.~E.,  {Gavazzi} R.,  {Marshall}
  P.~J.,   {Bolton} A.~S.,  2010, \mn@doi [\apj]
  {10.1088/0004-637X/709/2/1195}, \href
  {http://adsabs.harvard.edu/abs/2010ApJ...709.1195T} {709, 1195}

\bibitem[\protect\citeauthoryear{{Trujillo}, {Ferr{\'e}-Mateu}, {Balcells},
  {Vazdekis}  \& {S{\'a}nchez-Bl{\'a}zquez}}{{Trujillo}
  et~al.}{2014}]{2014ApJ...780L..20T}
{Trujillo} I.,  {Ferr{\'e}-Mateu} A.,  {Balcells} M.,  {Vazdekis} A.,
  {S{\'a}nchez-Bl{\'a}zquez} P.,  2014, \mn@doi [\apjl]
  {10.1088/2041-8205/780/2/L20}, \href
  {http://adsabs.harvard.edu/abs/2014ApJ...780L..20T} {780, L20}

\bibitem[\protect\citeauthoryear{{Van Der Walt}, {Colbert}  \&
  {Varoquaux}}{{Van Der Walt} et~al.}{2011}]{numpy}
{Van Der Walt} S.,  {Colbert} S.~C.,   {Varoquaux} G.,  2011, preprint, \href
  {http://adsabs.harvard.edu/abs/2011arXiv1102.1523V} {} (\mn@eprint {arXiv}
  {1102.1523})

\bibitem[\protect\citeauthoryear{{Vazdekis}, {Casuso}, {Peletier}  \&
  {Beckman}}{{Vazdekis} et~al.}{1996}]{1996ApJS..106..307V}
{Vazdekis} A.,  {Casuso} E.,  {Peletier} R.~F.,   {Beckman} J.~E.,  1996,
  \mn@doi [\apjs] {10.1086/192340}, \href
  {http://adsabs.harvard.edu/abs/1996ApJS..106..307V} {106, 307}

\bibitem[\protect\citeauthoryear{{Walsh}, {van den Bosch}, {Gebhardt},
  {Y{\i}ld{\i}r{\i}m}, {Richstone}, {G{\"u}ltekin}  \& {Husemann}}{{Walsh}
  et~al.}{2016}]{2016ApJ...817....2W}
{Walsh} J.~L.,  {van den Bosch} R.~C.~E.,  {Gebhardt} K.,  {Y{\i}ld{\i}r{\i}m}
  A.,  {Richstone} D.~O.,  {G{\"u}ltekin} K.,   {Husemann} B.,  2016, \mn@doi
  [\apj] {10.3847/0004-637X/817/1/2}, \href
  {http://adsabs.harvard.edu/abs/2016ApJ...817....2W} {817, 2}

\bibitem[\protect\citeauthoryear{Waskom et~al.,}{Waskom et~al.}{2015}]{seaborn}
Waskom M.,  et~al., 2015, seaborn: v0.6.0 (June 2015),
  \mn@doi{10.5281/zenodo.19108}, \url {https://doi.org/10.5281/zenodo.19108}

\bibitem[\protect\citeauthoryear{{Weijmans} et~al.,}{{Weijmans}
  et~al.}{2009}]{2009MNRAS.398..561W}
{Weijmans} A.-M.,  et~al., 2009, \mn@doi [\mnras]
  {10.1111/j.1365-2966.2009.15134.x}, \href
  {http://adsabs.harvard.edu/abs/2009MNRAS.398..561W} {398, 561}

\bibitem[\protect\citeauthoryear{{Wing} \& {Ford}}{{Wing} \&
  {Ford}}{1969}]{1969PASP...81..527W}
{Wing} R.~F.,  {Ford} Jr. W.~K.,  1969, \mn@doi [\pasp] {10.1086/128814}, \href
  {http://adsabs.harvard.edu/abs/1969PASP...81..527W} {81, 527}

\bibitem[\protect\citeauthoryear{{Worthey}, {Faber}, {Gonzalez}  \&
  {Burstein}}{{Worthey} et~al.}{1994}]{1994ApJS...94..687W}
{Worthey} G.,  {Faber} S.~M.,  {Gonzalez} J.~J.,   {Burstein} D.,  1994,
  \mn@doi [\apjs] {10.1086/192087}, \href
  {http://adsabs.harvard.edu/abs/1994ApJS...94..687W} {94, 687}

\bibitem[\protect\citeauthoryear{{Worthey}, {Tang}  \& {Serven}}{{Worthey}
  et~al.}{2014}]{2014ApJ...783...20W}
{Worthey} G.,  {Tang} B.,   {Serven} J.,  2014, \mn@doi [\apj]
  {10.1088/0004-637X/783/1/20}, \href
  {http://adsabs.harvard.edu/abs/2014ApJ...783...20W} {783, 20}

\bibitem[\protect\citeauthoryear{{Y{\i}ld{\i}r{\i}m}, {van den Bosch}, {van de
  Ven}, {Husemann}, {Lyubenova}, {Walsh}, {Gebhardt}  \&
  {G{\"u}ltekin}}{{Y{\i}ld{\i}r{\i}m} et~al.}{2015}]{2015MNRAS.452.1792Y}
{Y{\i}ld{\i}r{\i}m} A.,  {van den Bosch} R.~C.~E.,  {van de Ven} G.,
  {Husemann} B.,  {Lyubenova} M.,  {Walsh} J.~L.,  {Gebhardt} K.,
  {G{\"u}ltekin} K.,  2015, \mn@doi [\mnras] {10.1093/mnras/stv1381}, \href
  {http://adsabs.harvard.edu/abs/2015MNRAS.452.1792Y} {452, 1792}

\bibitem[\protect\citeauthoryear{{Zieleniewski}, {Houghton}, {Thatte}  \&
  {Davies}}{{Zieleniewski} et~al.}{2015}]{Z15}
{Zieleniewski} S.,  {Houghton} R.~C.~W.,  {Thatte} N.,   {Davies} R.~L.,  2015,
  \mn@doi [\mnras] {10.1093/mnras/stv1251}, \href
  {http://adsabs.harvard.edu/abs/2015MNRAS.452..597Z} {452, 597}

\bibitem[\protect\citeauthoryear{{Zieleniewski}, {Houghton}, {Thatte}, {Davies}
   \& {Vaughan}}{{Zieleniewski} et~al.}{2017}]{2017MNRAS.465..192Z}
{Zieleniewski} S.,  {Houghton} R.~C.~W.,  {Thatte} N.,  {Davies} R.~L.,
  {Vaughan} S.~P.,  2017, \mn@doi [\mnras] {10.1093/mnras/stw2712}, \href
  {http://adsabs.harvard.edu/abs/2017MNRAS.465..192Z} {465, 192}

\bibitem[\protect\citeauthoryear{{van Dokkum}}{{van Dokkum}}{2001}]{LaCosmic}
{van Dokkum} P.~G.,  2001, \mn@doi [\pasp] {10.1086/323894}, \href
  {http://adsabs.harvard.edu/abs/2001PASP..113.1420V} {113, 1420}

\bibitem[\protect\citeauthoryear{{van Dokkum} \& {Conroy}}{{van Dokkum} \&
  {Conroy}}{2010}]{2010Natur.468..940V}
{van Dokkum} P.~G.,  {Conroy} C.,  2010, \mn@doi [\nat] {10.1038/nature09578},
  \href {http://adsabs.harvard.edu/abs/2010Natur.468..940V} {468, 940}

\bibitem[\protect\citeauthoryear{{van Dokkum} \& {Conroy}}{{van Dokkum} \&
  {Conroy}}{2012}]{CvD12}
{van Dokkum} P.~G.,  {Conroy} C.,  2012, \mn@doi [\apj]
  {10.1088/0004-637X/760/1/70}, \href
  {http://adsabs.harvard.edu/abs/2012ApJ...760...70V} {760, 70}

\bibitem[\protect\citeauthoryear{{van Dokkum}, {Conroy}, {Villaume}, {Brodie}
  \& {Romanowsky}}{{van Dokkum} et~al.}{2017}]{2017ApJ...841...68V}
{van Dokkum} P.,  {Conroy} C.,  {Villaume} A.,  {Brodie} J.,   {Romanowsky}
  A.~J.,  2017, \mn@doi [\apj] {10.3847/1538-4357/aa7135}, \href
  {http://adsabs.harvard.edu/abs/2017ApJ...841...68V} {841, 68}

\bibitem[\protect\citeauthoryear{{van den Bosch}, {Gebhardt}, {G{\"u}ltekin},
  {van de Ven}, {van der Wel}  \& {Walsh}}{{van den Bosch}
  et~al.}{2012}]{2012Natur.491..729V}
{van den Bosch} R.~C.~E.,  {Gebhardt} K.,  {G{\"u}ltekin} K.,  {van de Ven} G.,
   {van der Wel} A.,   {Walsh} J.~L.,  2012, \mn@doi [\nat]
  {10.1038/nature11592}, \href
  {http://adsabs.harvard.edu/abs/2012Natur.491..729V} {491, 729}

\makeatother
\end{thebibliography}

\appendix

\section{Index Measurements}

We present our radial index measurements in Tables \ref{tab:all_inds_NGC1277} and \ref{tab:all_inds_IC843}. The methodology behind these measurements is discussed in Section \ref{sec:IndexCorrectionFactors}.

\begin{table*}
\caption{All index measurements in NGC~1277. As discussed in Section \ref{sec:IndexCorrectionFactors}, these measurements were taken at the intrinsic velocity dispersion of the radial bin and then corrected to 200 \kms~for both galaxies. All results are equivalent widths, in units of \AA~and found using the formalism of \citet{Cenarro2001}, except for that of TiO which is a ratio of the blue and red pseudo-continuua. }
\begin{center}
\begin{tabular}{cccccc}
\hline
$\log(R/R_{e})$ & CaT (\AA)&                  FeH (\AA)&                   MgI (\AA)&                   NaI (\AA)&                   TiO\\
\hline
-0.99 &$6.94\pm0.34$ & $0.46\pm0.13$ & $0.36\pm0.16$ & $1.28\pm0.12$ & $1.0809\pm0.0062$\\ 
-0.60 &$6.78\pm0.29$ & $0.48\pm0.11$ & $0.25\pm0.12$ & $1.20\pm0.10$ & $1.0795\pm0.0053$\\ 
-0.40 &$6.71\pm0.40$ & $0.41\pm0.15$ & $0.33\pm0.15$ & $1.10\pm0.13$ & $1.0788\pm0.0069$\\ 
-0.21 &$6.73\pm0.40$ & $0.44\pm0.15$ & $0.42\pm0.15$ & $1.04\pm0.14$ & $1.0764\pm0.0070$\\ 
-0.06 &$6.61\pm0.32$ & $0.51\pm0.12$ & $0.68\pm0.11$ & $1.02\pm0.11$ & $1.0754\pm0.0057$\\ 
0.11 &$6.18\pm0.36$ & $0.62\pm0.12$ & $0.57\pm0.12$ & $0.97\pm0.12$ & $1.0679\pm0.0061$\\ 
0.28 &$6.11\pm0.31$ & $0.50\pm0.10$ & $0.57\pm0.11$ & $0.75\pm0.11$ & $1.0622\pm0.0052$\\ 
Global &$6.82\pm0.13$ & $0.45\pm0.05$ & $0.56\pm0.07$ & $1.07\pm0.05$ & $1.0729\pm0.0030$\\ 
\end{tabular}
\end{center}
\label{tab:all_inds_NGC1277}
\end{table*}

\begin{table*}
\caption{All index measurements in IC~843. See the caption of Table \ref{tab:all_inds_NGC1277} for details}
\begin{center}
\begin{tabular}{cccccc}
\hline
$\log(R/R_{e})$ & CaT (\AA)&                  FeH (\AA)&                   MgI (\AA)&                   NaI (\AA)&                   TiO\\
\hline
-1.60 &$7.34\pm0.40$ & $0.42\pm0.16$ & $0.58\pm0.13$ & $0.87\pm0.15$ & $1.0718\pm0.0070$\\ 
-1.06 &$7.28\pm0.33$ & $0.50\pm0.13$ & $0.54\pm0.10$ & $0.74\pm0.12$ & $1.0686\pm0.0057$\\ 
-0.79 &$7.17\pm0.33$ & $0.53\pm0.12$ & $0.57\pm0.10$ & $0.68\pm0.12$ & $1.0683\pm0.0056$\\ 
-0.60 &$7.36\pm0.35$ & $0.43\pm0.13$ & $0.58\pm0.09$ & $0.64\pm0.13$ & $1.0663\pm0.0058$\\ 
-0.47 &$7.24\pm0.37$ & $0.41\pm0.13$ & $0.54\pm0.09$ & $0.57\pm0.13$ & $1.0647\pm0.0060$\\ 
-0.35 &$7.20\pm0.41$ & $0.39\pm0.14$ & $0.54\pm0.10$ & $0.55\pm0.14$ & $1.0661\pm0.0064$\\ 
-0.24 &$6.97\pm0.37$ & $0.45\pm0.13$ & $0.58\pm0.09$ & $0.55\pm0.13$ & $1.0621\pm0.0057$\\ 
Global &$7.24\pm0.14$ & $0.41\pm0.05$ & $0.58\pm0.04$ & $0.66\pm0.05$ & $1.0671\pm0.0024$\\ 
\end{tabular}
\end{center}
\label{tab:all_inds_IC843}
\end{table*}

\section{Sky Subtraction Methods}
\label{Appendix:SkySub}
\subsection{Sky subtraction with \texttt{pPXF}}
First order sky subtraction was applied to each galaxy cube. This involved subtracting a separate "sky" cube, made by combining sky observations taken throughout the night, from the combined galaxy data. Since night sky emission lines vary on timescales of minutes, similar to the length of our observations, such a first order sky subtraction will not be perfect and the resulting (sky subtracted) spectra still contain residual sky light. To correct for this residual sky emission, we use \ppxf~to fit a set of sky templates at the same time as measuring the kinematics from each spectrum, a process first described in \citet{2009MNRAS.398..561W}. 

\begin{figure*}
\centering
\includegraphics[width=\linewidth, height=10cm]{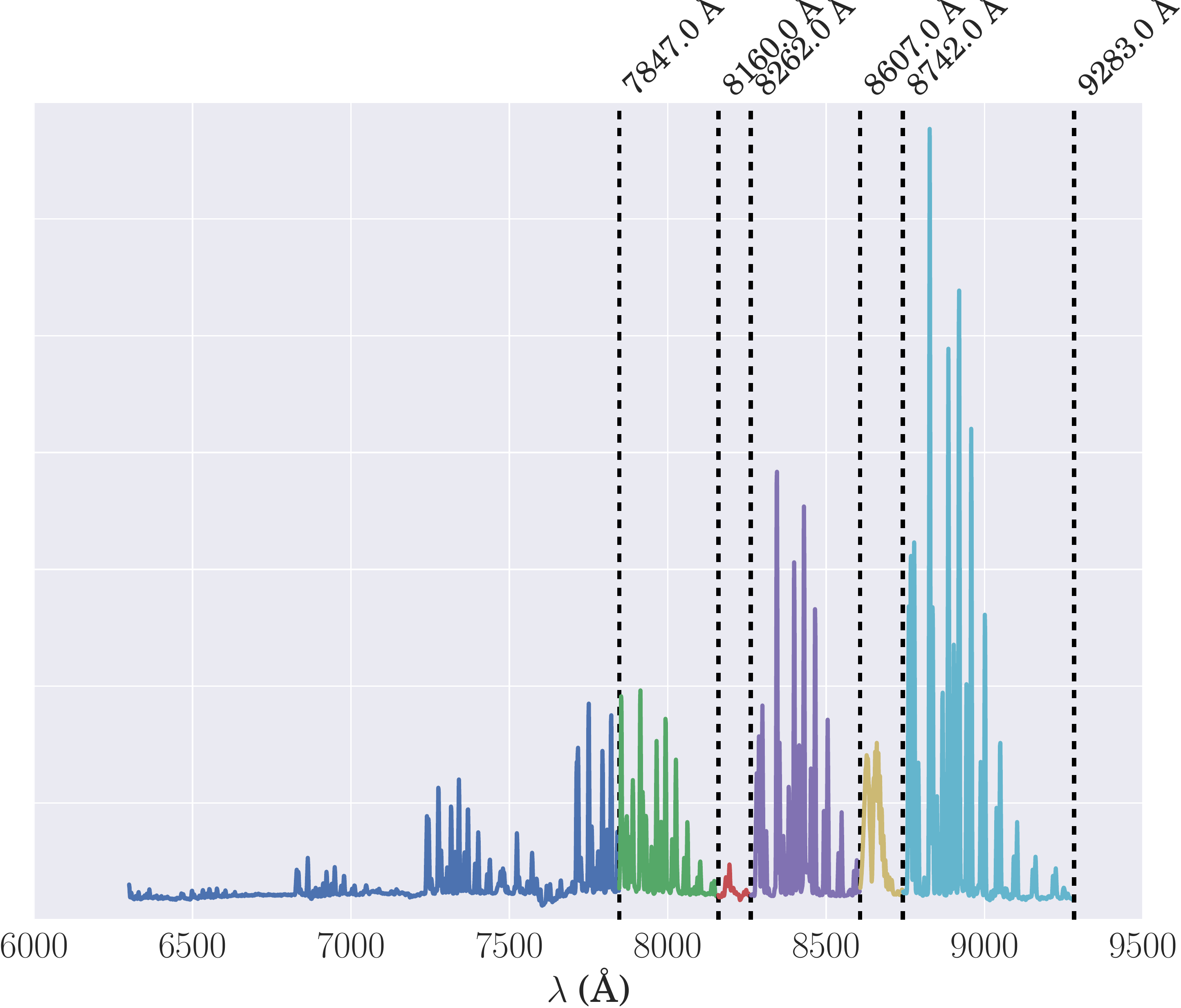}
\caption{A sky spectrum from 6300 \AA~to 9283 \AA, showing locations of split locations.}
\label{fig:Sky_1}

\end{figure*}

\begin{figure*}
\centering
\includegraphics[width=\linewidth, height=10cm]{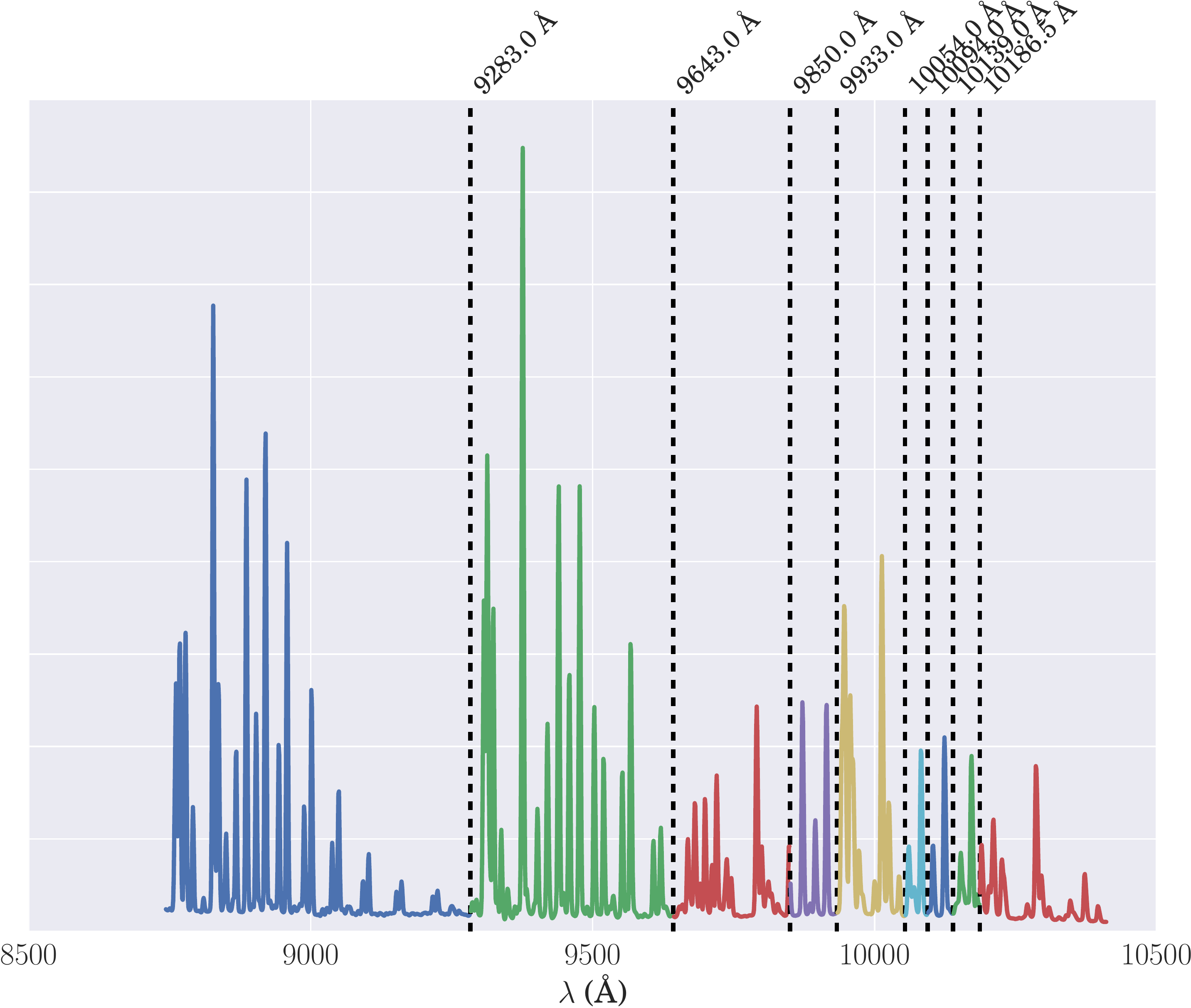}
\caption{A sky spectrum centred around the Wing Ford band, showing the locations  of sky splits. At the redshift of NGC~1277 and IC~843, the Wing-Ford band is observed at 10085 \AA~and 10160 \AA~respectively.}
\label{fig:Sky_FeH}

\end{figure*}

The sky templates are made from the sky cube used for first order sky subtraction. The observed sky spectrum is split around selected molecular bandheads and transitions, according to wavelengths defined in \cite{2007MNRAS.375.1099D}, so that emission lines corresponding to different molecules are allowed to be scaled separately in \ppxf. We introduced a small number of further splits, around where sky residuals were seen to sharply change sign. We also allow for over-subtracted skylines by including negatively scaled sky spectra, and account for instrument flexure by including sky spectra which have been shifted forwards and backwards in wavelength by up to 2.5 \AA~(in 0.5 \AA~increments). A full sky spectrum, with locations of sky splits marked, is shown in Figures \ref{fig:Sky_1} and \ref{fig:Sky_FeH}.

The area worst affected by residual sky emission is the Wing-Ford band at 9916 \AA. Here, we found that changing the combination of sky splits had an impact on the quality of sky subtraction, and hence on the FeH equivalent width measurement. To quantitatively choose the set of skyline splits which gave us the best sky subtraction, we investigated the residuals of the sky subtracted spectrum around the best fitting \ppxf~template. These residuals will generally be distributed like a Gaussian around zero, with any remaining skyline residuals appearing as large positive or negative outliers. A set of residuals which have tails which deviate from a normal distribution therefore imply a poor sky subtraction.  

\begin{figure}
\centering
\includegraphics[width=\linewidth]{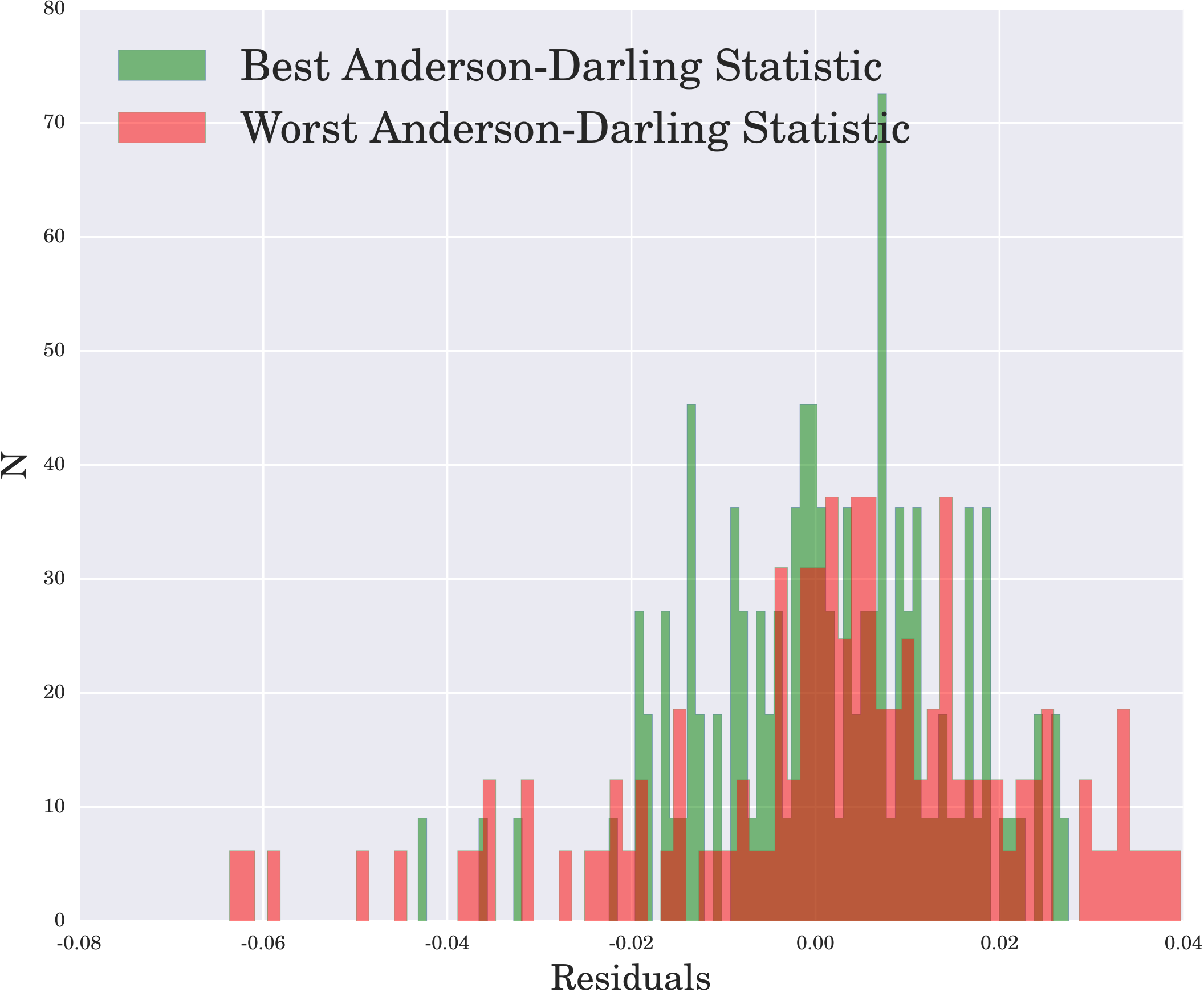}
\caption{Residuals around the best fit \ppxf~template, for one of the NGC~1277 outer bins, after second order sky subtraction using \ppxf.  The two histograms correspond to two different combinations of sky line splits. The histogram in green shows the distribution with the lowest Anderson-Darling test statistic of all 32 sky split combinations, whilst the one in red shows a distribution with many outlying residuals and a large A-D statistic, corresponding to a poor second-order sky subtraction. }
\label{fig:residual_histograms}

\end{figure}

\begin{figure*}
\centering
\includegraphics[width=\linewidth]{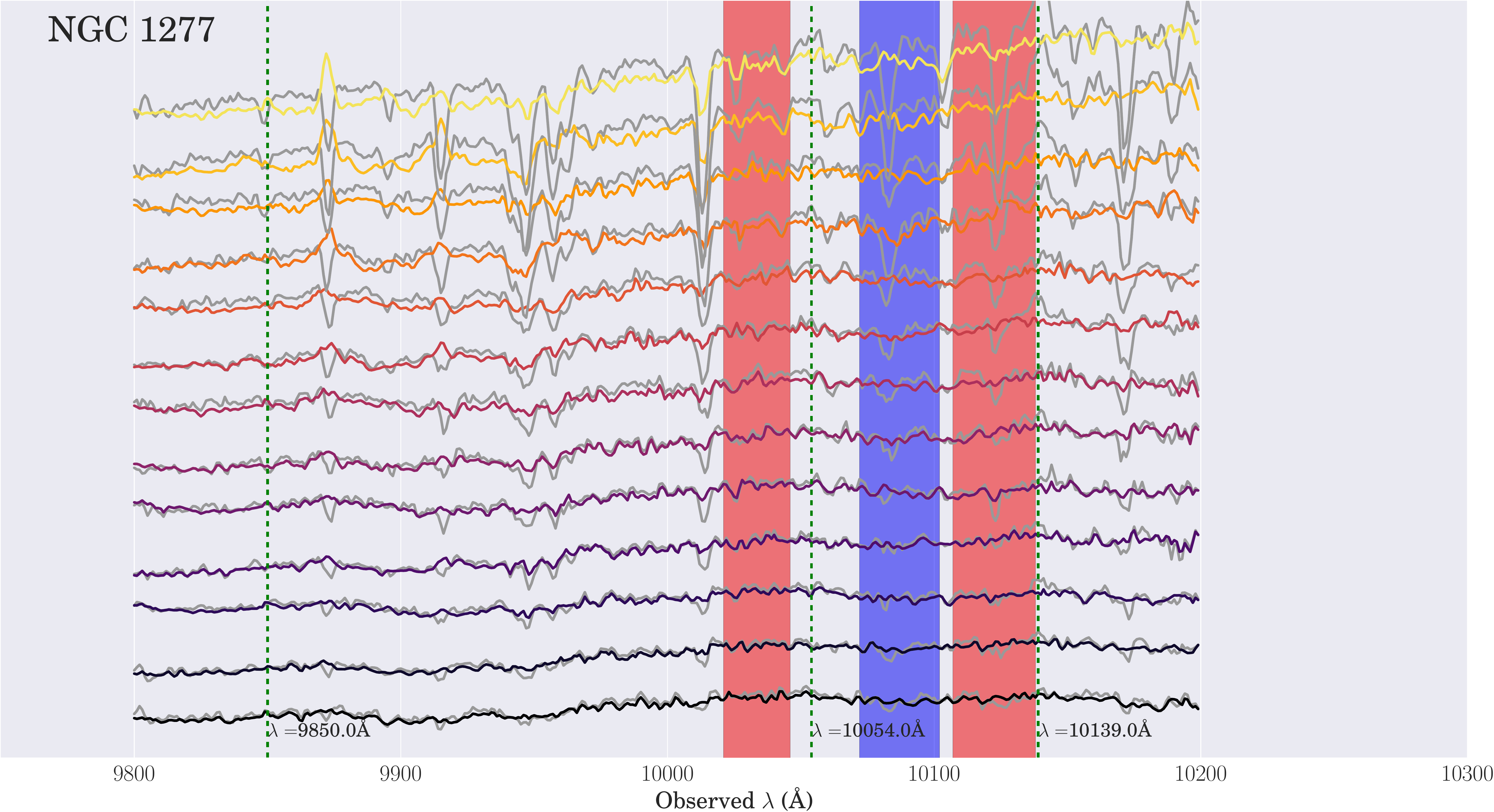}
\includegraphics[width=\linewidth]{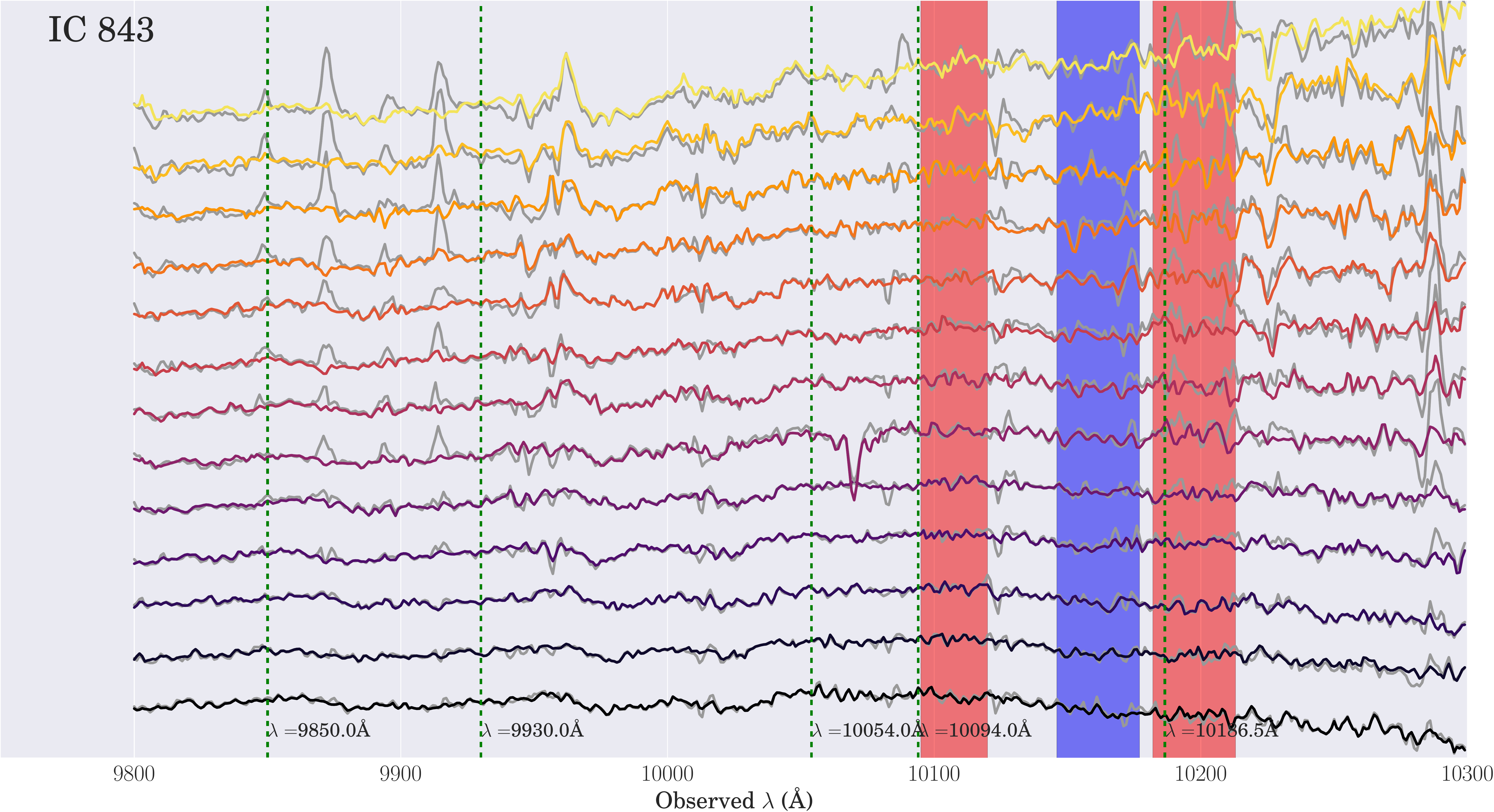}
\caption{13 spectra around the Wing-Ford band for both NGC~1277  (top) and IC~843 (bottom). These 13 spectra correspond to a central bin and 6 spectra each from the left and right sides of the galaxy. The 7 radial bins for each galaxy we use to measure index equivalent widths are formed by interpolating each of these spectra to their rest frame and adding them. The lines correspond to the spectrum in each bin before \ppxf~sky subtraction (but after first order sky subtraction; coloured grey) and afterwards (black through yellow). The spectra range from the central bin (bottom, dark) to the outermost (top, light). Green dashed line indicate the position of a cut to the sky spectrum. Blue shaded regions show the location of the FeH index whilst red shaded regions identify the location of the continuum regions.  }
\label{fig:ppxf_sky_sub}

\end{figure*}

Around FeH, there are 5 wavelengths which we decided to split the sky at; 9933 \AA, 10054 \AA, 10094 \AA, 1013 9\AA~and 1.01865 \AA.  This leads to $2^5=32$ possible combinations of splits. We investigated the residuals for each of these 32 combinations, both by eye and using an Anderson-Darling test \citep[][AD]{AndersonDarling} with the null hypothesis that each sample was drawn from a normal distribution. The AD test is very similar to the more commonly used Kolmogorov-Smirnov test, except with a weighting function which emphasises the tails of each distribution more than a KS test does. For all analysis in this work we used the selection of sky splits with the lowest AD statistic, which corresponds to the residual distribution best described by a normal distribution with no outliers. A plot of the best (green) and worst (red) residual distribution for the NGC~1277 skylines is shown in Figure \ref{fig:residual_histograms}, whilst Figure \ref{fig:ppxf_sky_sub} shows our spectra around FeH for NGC~1277 and IC~843 before and after second order sky subtraction. The spectra which, by eye, have the best sky subtraction are also those with the lowest AD statistic.

\subsection{A comparison of independent sky subtraction methods}
\label{Appendix:sky_sub_comp}

\begin{figure*}
\centering
\includegraphics[width=\linewidth]{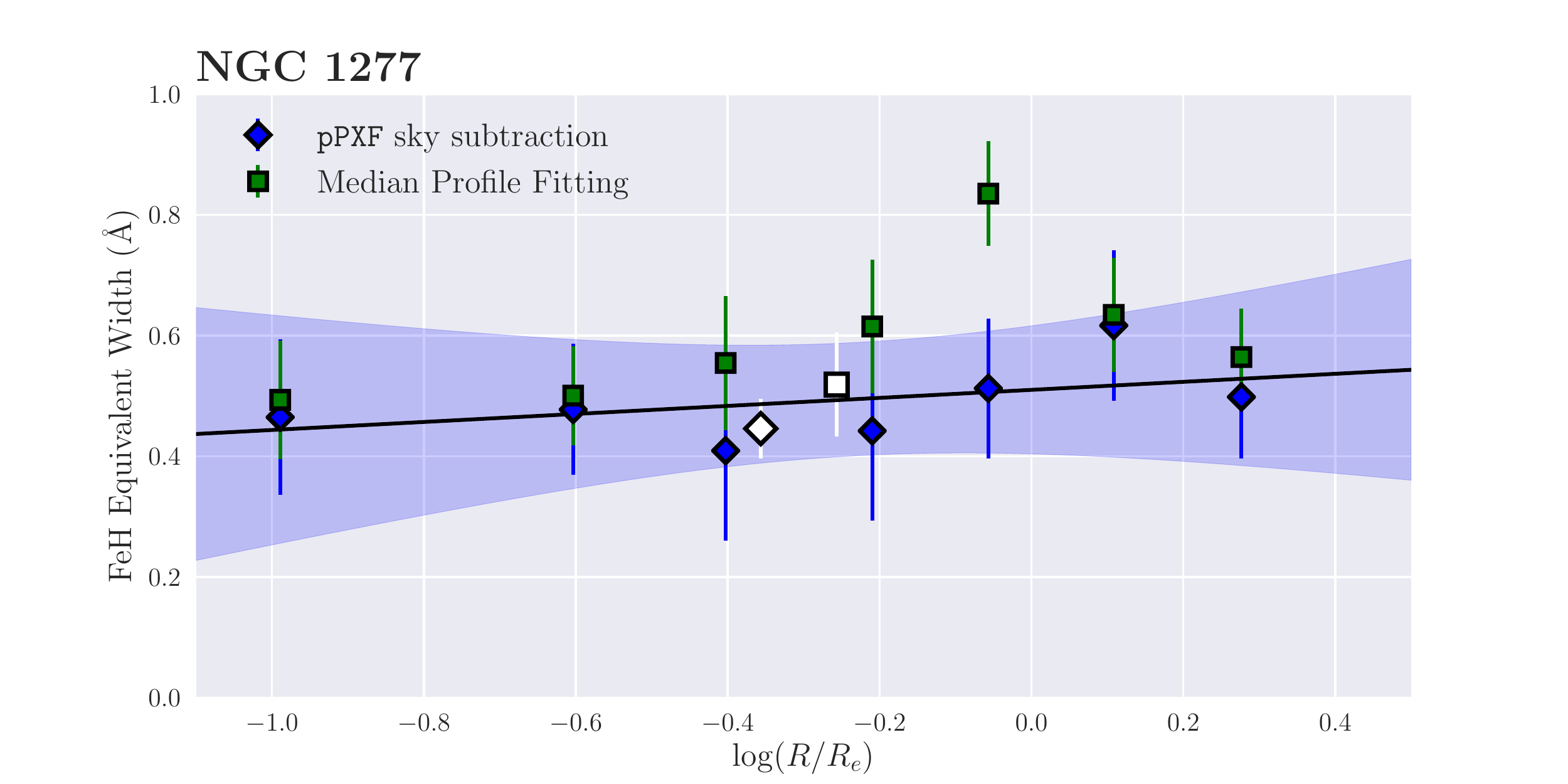}
\includegraphics[width=\linewidth]{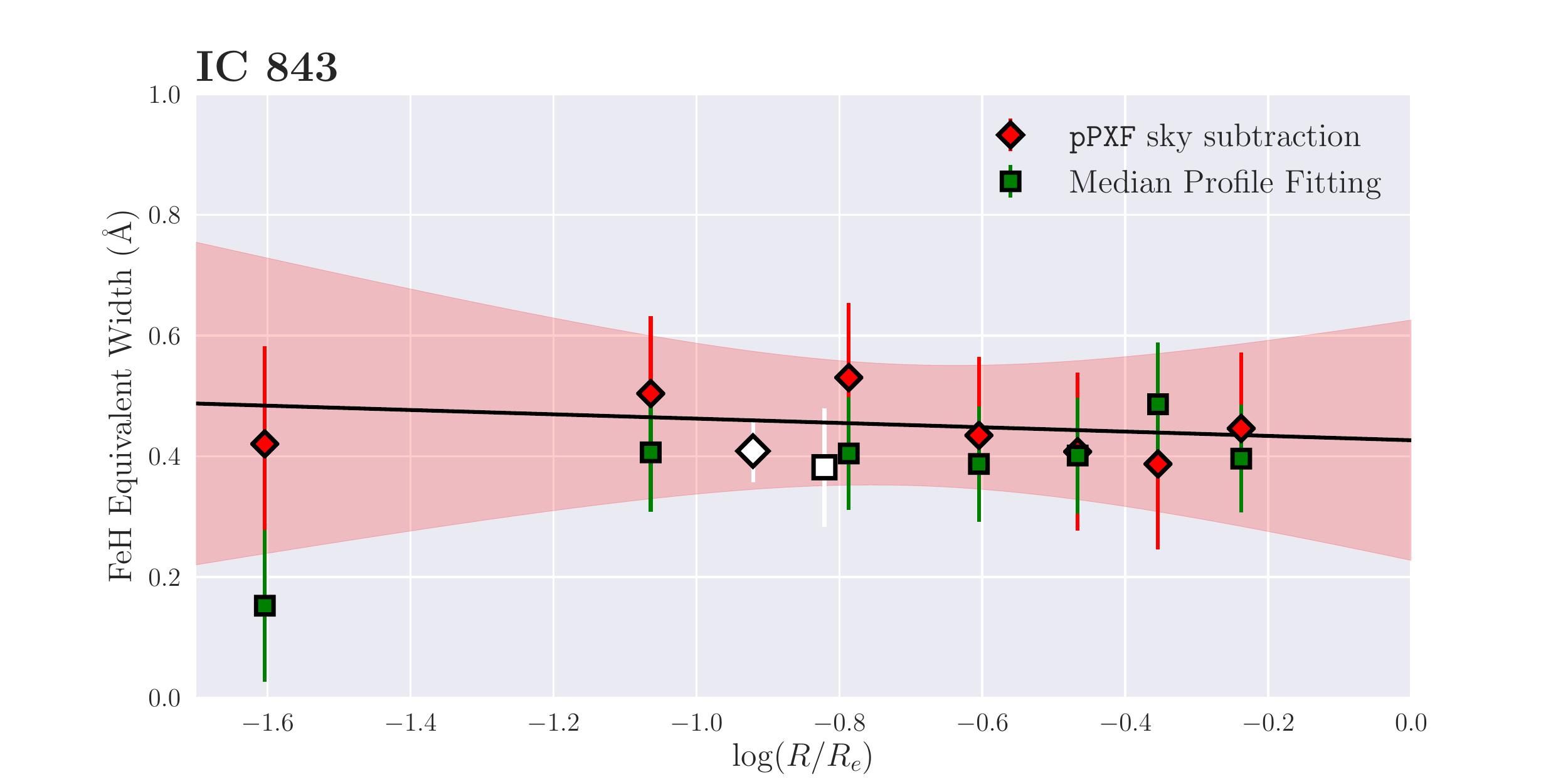}
\caption{A comparison of equivalent width measurements of the Wing-Ford band, after both \ppxf~sky subtraction and median profile fitting, for NGC~1277 (top) and IC~843 (bottom). The best fit straight line for the \ppxf~sky subtraction is shown for both galaxies, as well as its one sigma uncertainty (shaded region). The two sky subtraction methods give generally good agreement, confirming the robustness of our results. In particular, the global FeH measurements are entirely consistent between the two approaches.}
\label{fig:sky_sub_comparison}

\end{figure*}

Figure \ref{fig:sky_sub_comparison} shows equivalent width measurements of the Wing Ford band, found using spectra from the two sky subtraction processes; subtracting skylines with \ppxf~and median profile fitting. These methods are described in Section \ref{sec:sky_subtraction} and Appendix \ref{Appendix:SkySub}.

The two methods show good agreement, implying that our FeH measurements are robust, despite the challenging nature of removing residual sky emission in the far red region of the spectrum. In particular, the global FeH measurements, on which we base our determination of the IMF in these galaxies, are entirely consistent between the two approaches.

\section{Comparison to stellar population models}
\label{Appendix:IMF_Calculations}

In order to make quantitative measurements of the IMF in each galaxy, we compare our results to the CvD12 and E-MILES stellar population models. The aim is to create a spectrum with the same stellar population parameters and FeH measurement as the global spectrum for both galaxies, and then read off the IMF slope of that spectrum. The two sets of models allow for changes in separate population parameters, meaning that the analysis which starts with a base template from the CvD12 models is slightly different to the case where we start with a base spectrum from the E-MILES models. Both cases are described below.

\subsection{CvD12}

We interpolate the base set of CvD12 models of varying IMF slope as a function of age and FeH equivalent width. These base spectra are at solar metallicity, [$\alpha$/Fe]=0 and have solar elemental abundance ratios, whilst spanning IMF slopes from bottom-light to $x=3.5$. To accurately account for the different metallicities, $\alpha$-abundances and [Na/Fe] ratios in each galaxy, we apply linear response functions to the CvD spectra.

The correction is defined as follows. To deal with varying continuum levels between spectra with different IMF slopes, we use multiplicative rather than additive response functions. For a spectrum with a non-solar $\alpha$-abundance ratio, $S(\Delta \alpha)$, 

\begin{equation*}
S(\Delta \alpha) = x_{\alpha} S(\Delta \alpha=0.0)
\end{equation*}

\noindent where $x_{\alpha}$ is the linear response function. We also Taylor expand $S(\Delta \alpha)$ to give

\begin{equation*}
S(\Delta \alpha) \approx S(\Delta \alpha=0.0)+\frac{dS}{d\alpha}\Delta \alpha
\end{equation*}

\noindent which leaves

\begin{equation*}
\centering
x_{\alpha} =(1+\frac{d\ln S}{d\alpha}\Delta \alpha)
\end{equation*}

\noindent We approximate the gradient term using a model spectrum from CvD12 at enhanced [$\alpha$/Fe]=+0.3:

\begin{align*}
\frac{d\ln S}{d\alpha}\Delta \alpha &\approx \frac{1}{S(\Delta \alpha = +0.0)}\frac{S(\Delta \alpha = +0.3)-S(\Delta \alpha = +0.0)}{10^{0.3}-1}(10^{\Delta \alpha}-1) \\
 &=\left(\frac{S(\Delta \alpha = +0.3)}{S(\Delta \alpha = +0.0)}-1\right)\frac{10^{\Delta \alpha}-1}{10^{0.3}-1}\\
 &=f_{\alpha}
\end{align*}

A similar correction is applied for [Fe/H] and [Na/Fe] abundance variations.

The final set of spectra are therefore:

\begin{align*}
S_{\text{final}} &= S_0\cdot x_{\alpha}\cdot x_{\text{Na}}\cdot x_{\text{Fe}}\\
\ln S_{\text{final}}& = \ln S_{0} + f_{\alpha} +f_{\text{Na}} + f_{\text{Fe}}
\end{align*}

It is important to note that the CvD12 spectra with non-elemental abundances (e.g those with [Fe/H]=+0.3 dex) are calculated from a Chabrier IMF, whereas we find the IMFs in these galaxies from this analysis to be heavier than this. Another unavoidable source of uncertainty in the use of these models concerns the fact that the response of the IMF sensitive indices strong in very low-mass stars (such as FeH) to quantities like [$\alpha$/Fe] are computed from theoretical atmospheric models which may not converge. For further discussion of this point, see CvD12 section 2.4.

\subsection{E-MILES}

A similar process was carried out for the E-MILES spectra. We interpolate a grid of templates of varying IMF, age, metallicity and [Na/Fe] enhancement. Since the E-MILES models are all at solar [$\alpha$/Fe] abundance, we use a response function from the CvD12 models to approximate an $\alpha$-enhanced spectrum. 

A complication here is that a CvD12 model template at [$\alpha$/Fe]=+0.3 is not at solar metallicity, because the CvD12 models are computed at fixed [Fe/H] and not fixed [Z/H]. Using the relation from \cite{2000AJ....120..165T},  

\begin{equation*}
\label{eqtn:CvD12_metallicity}
\textrm{[Fe/H]} = \textrm{[Z/H]}-0.93\times\textrm{[$\alpha$/Fe]}
\end{equation*}
\smallskip

\noindent and so CvD12 template with [$\alpha$/Fe]=+0.3 also has [Z/H]=0.279. We must therefore apply an [$\alpha$/Fe] response function to a base spectrum of metallicity 

\begin{equation*}
\textrm{[Z/H]}_\textrm{spectrum}=\textrm{[Z/H]}_\textrm{galaxy}-0.93\times\textrm{[$\alpha$/Fe]}. 
\end{equation*}
\smallskip

 \noindent rather than simply $\textrm{[Z/H]}_\textrm{galaxy}$. This means, therefore, that the base template used for NGC~1277 has [Z/H]=-0.079, whilst the base template for IC~843 has [Z/H]=-0.197.

\bsp	\label{lastpage}
\end{document}